\documentclass[letterpaper,prc,preprint]{revtex4}
\usepackage[dvips]{graphicx}
\usepackage{epsfig}
\usepackage{amssymb, latexsym, amsmath, amsfonts, fullpage}

\newcommand{\be}{\begin{equation}}
\newcommand{\ee}{\end{equation}}
\newcommand{\bea}{\begin{eqnarray}}
\newcommand{\eea}{\end{eqnarray}}

\begin{document}

\title{Natural orbitals representation and Fermi sea depletion in finite nuclei and nuclear matter}

\author{V.P. Psonis\footnote{bpson@auth.gr}, Ch.C. Moustakidis\footnote{moustaki@auth.gr}, and S.E. Massen\footnote{massen@auth.gr},}

\address{Department of Theoretical Physics, Aristotle University of
Thessaloniki,  54124 Thessaloniki, Greece}

\begin{abstract}
The natural orbitals and natural occupation numbers of various
$N=Z$, $sp$ and $sd$ shell nuclei are calculated by applying a
correlated one-body density matrix. The correlated density matrix
has been evaluated by considering  central correlations of Jastrow
type and an approximation named factor cluster expansion. The
correlation effects on the natural orbitals, natural occupation
numbers and the Fermi sea depletion are discussed and analysed. In
addition, an approximate expression for the correlated one-body
density matrix of the nuclear matter has been used for the
evaluation of the relative momentum distribution and the Fermi sea
depletion. We found that the value of the Fermi sea depletion is higher in
closed shell nuclei compared to open shell ones and it is
lower compared to the case of nuclear matter. This statement could
be confirmed by relevant experimental studies.

\vspace{0.3cm}

PACS number(s): 21.10.Gv, 21.60.-n, 21.60.Cs, 21.65.-f.
 \\

\end{abstract}

\maketitle
\newpage

\section{Introduction}

The short-range correlations (SRC) play an important role on the
one- and two-body properties of nuclei and nuclear matter. In
general, the contribution of SRC is important for the description
of the mean value of some two-body operators, such as the
ground-state energy of nuclei, but it is also of interest to
investigate the SRC contribution to simpler nuclear quantities
related to one-body operators such as the form factor, density
distribution and momentum distribution. The key quantity to
calculate the one-body properties of a quantum many-body system
is the one-body density matrix (OBDM)\cite{Lowdin-55}. The
knowledge of the OBDM leads to the evaluation of the one-body
properties of a quantum many-body system both in coordinate and in
momentum space. In addition, the diagonalization of the OBDM leads
to the evaluation of the natural orbitals (NO) and natural
occupation numbers (NON). Over the last decades there was an
effort to incorporate SRC on the OBDM mainly for light and medium
nuclei~\cite{Bohigas-80,Jaminon-85,Mahaux-87,Arias-97,
Stoitsov-93,Massen-99,Moustakidis-00,Moustakidis-02-1,
Moustakidis-01,Chatzissavas-05,Moustakidis-07,
Papakonstantinou-03,Mavrommatis-01,Bisconti-07,Lacroix-09,Dickhoff-10}.

The OBDM has been evaluated  recently, by considering  central
correlations of Jastrow type \cite{Jastrow-55} in the framework of
the Iwamoto and Yamada factor (FIY) cluster
expansion~\cite{Iwamoto-57} developed by Clark and Westhaus
\cite{Clark-67} and Feenberg \cite{Feenberg-69}. More precisely,
the expression of the OBDM, $\rho({\bf r},{\bf r}')$, was found by
using the factor cluster expansion of Clark and co-workers
\cite{Clark-79} and Jastrow correlation function which introduces
SRC for $N=Z$ $sp$ and $sd$ shell nuclei. The evaluated OBDM is a
functional of the harmonic oscillator (HO) orbitals and depends on
the HO parameter $b=(\hbar/m\omega)^{1/2}$ and the correlation
parameter $\beta$ \cite{Moustakidis-00}.

In addition,  by employing a phenomenological ansantz for the
correlated OBDM of the uniform nuclear matter we calculated the
relative  momentum distribution and  consequently the Fermi sea
depletion (FSD). We considered a value for the nuclear matter density
close to equilibrium density of symmetrical nuclear matter in
order to be able to compare  the correlated parameter and consequently the effect of
SRC in finite nuclei and in nuclear matter.

The motivation of the present work is twofold. Firstly, we applied
the OBDM corresponding to finite nuclei in order to make a
detailed investigation of the correlation effects on the NO, the
natural occupation probabilities and the depletion of the Fermi
sea in light and medium nuclei. Actually, very little information
can be found in physics literature about the properties of natural
orbitals and occupation probabilities. In view of the above
statement,  we additionally extend the study of the NO and NON not
only to closed shell  but also to some open shell nuclei. To
our knowledge this is an intriguing study that has never been
examined. Secondly, we extended the calculations in nuclear matter  in order
to compare the effect of the SRC in finite and infinite nuclear
matter under almost the same conditions and to provide
results which could pose for future experimental investigation.

The article is organized as follows. In Sec.~II the correlated OBDM and
the diagonalization process in finite
nuclei are presented, while the relative theory of nuclear matter is reported
in Sec.~III. The results are discussed  in Sec.~IV. Finally the summary of the work
is given in Sec.~V.

\section{Correlated one-body density  matrix and natural orbitals representation}
\subsection{One-body density matrix}
A nucleus with $A$ nucleons is described by the wave function
$\Psi({\bf r}_1,{\bf r}_2,\cdots,{\bf r}_A)$ which depends on $3A$
coordinates as well as on spins and isospins. The evaluation of
the single particle characteristic of the system requires the
one-body density matrix~\cite{Lowdin-55}
\begin{equation}
\rho({\bf r},{\bf r}')=\int\Psi^*({\bf r},{\bf r}_2,\cdots,{\bf
r}_A) \times \Psi({\bf r}',{\bf r}_2,\cdots,{\bf r}_A) d{\bf
r}_2 \cdots   d{\bf r}_A, \label{OBDM-1}
\end{equation}
where the integration is carried out over the radius vectors ${\bf
r}_2,\cdots,{\bf r}_A $ and summation over spin and isospin
variables is implied. $\rho({\bf r},{\bf r}')$ can also be
represented by the form
\begin{equation}
\rho({\bf r},{\bf r}')=\frac{\langle \Psi\mid {\bf O}_{{\bf r}{\bf
r}'}(A)\mid \Psi'\rangle}{\langle \Psi\mid\Psi \rangle}= N\langle
\Psi\mid {\bf O}_{{\bf r}{\bf r}'}(A)\mid \Psi'\rangle = N\langle
{\bf O}_{{\bf r}{\bf r}'}(A)\rangle, \label{OBDM-2}
\end{equation}
where $\Psi'=\Psi({\bf r}_1',{\bf r}_2',\cdots,{\bf r}_A')$ and $
N$ is the normalization factor. The one-body density operator
${\bf O}_{{\bf r}{\bf r}'}(A)$, has the form
\begin{equation}
{\bf O}_{{\bf r}{\bf r}'}(A)=\sum_{i=1}^A \delta({\bf r}_i-{\bf
r}) \delta({\bf r}_i'-{\bf r}')\prod_{j \neq i}^A \delta({\bf
r}_j-{\bf r}_j'). \label{OB-oper-1}
\end{equation}

The diagonal elements of the OBDM give the density distribution
$\rho({\bf r},{\bf r})=\rho({\bf r})$, while the momentum
distribution is given by the Fourier transform of $\rho({\bf
r},{\bf r}')$,
\begin{equation}
n({\bf k})=\frac{1}{(2\pi)^3}\int \exp[i{\bf k}({\bf r}-{\bf
r}')]\rho({\bf r},{\bf r}') d{\bf r}  d{\bf r}'.
\label{MD-1}
\end{equation}

The trial wave function $\Psi$, which describes a correlated
nuclear system, can be written as
\begin{equation}
\Psi={\cal F}\Phi, \label{Psi-1}
\end{equation}
where $\Phi$ is a model wave function that is adequate to describe
the uncorrelated $A$-particle nuclear system and ${\cal F}$ is the
operator that introduces SRC. The function $\Phi $, is chosen to
be  a Slater determinant  wave function, constructed by  $A$
ortho-normalised single-particle wave functions ${\phi_i({\bf
r})}$
\begin{equation}
\Phi=\Phi_{SD}({\bf r}_1,{\bf r}_2,\cdots,{\bf r}_A) =(A!)^{-1/2}
\det \mid \phi_i({\bf r}_j) \mid, \qquad (i,j=1,2,\cdots,A).
\label{Fi-SD}
\end{equation}
Several restrictions can be applied on the model operator ${\cal
F}$. In the present work ${\cal F}$ is taken to be of Jastrow
type~\cite{Jastrow-55}
\begin{equation}
{\cal F}=\prod_{i<j}^Af(r_{ij}), \label{F-1-1}
\end{equation}
where $f(r_{ij})$ is the state-independent correlation function of
the form
\begin{equation}
f(r_{ij})=1-\exp[-\beta({\bf r}_i-{\bf r}_j)^2]. \label{frij-1}
\end{equation}
The correlation function $f(r_{ij})$ tends to $1$ for large values
of $r_{ij}=|{\bf r}_i-{\bf r}_j|$ and it tends to $0$ for
$r_{ij}\rightarrow 1$.

The expression of the correlated one-body density matrix
$\rho({\bf r},{\bf r}')$ has been found by applying the method of
the generalised integrals and by using the factor cluster
expansion of Ristig, Ter Low, and Clark~\cite{Ristig-71}. The
various terms of $\rho({\bf r},{\bf r}')$ are given in
Ref.~\cite{Moustakidis-00}. The relative expression depends on the
single-particle wave functions and so it is suitable to be used
for analytical calculations with the HO orbitals and in principle
for numerical calculations with more realistic single-particle
orbitals. The calculations were carried out for $sp$ and $sd$ closed
shell and open shell nuclei with $N=Z$.

\subsection{Mean-field approximation and harmonic oscillator wave functions}
In the mean field approximation (MFA) the single-particle wave
functions $\phi_{\alpha}({\bf r})$ appearing in the A-body wave
function, given by Eq.~(\ref{Fi-SD}), form a nuclear Fermi sea F.
These wave functions are called hole-state orbitals and define the
uncorrelated OBDM
\begin{equation}
\rho_{MFA}({\bf r},{\bf r}')=\sum_{\alpha \in
F}\phi_{\alpha}^*({\bf r}) \phi_{\alpha}({\bf r}').
\label{MFA-OBDM}
\end{equation}
Eq.~(\ref{MFA-OBDM}) is associated  with the Slater determinant
wave function given by Eq.~(\ref{Fi-SD}). The uncorrelated density
matrix $\rho_{MFA}({\bf r},{\bf r}')$ is diagonal and
the occupation probabilities are equal to unity
inside the Fermi sea and zero for the states (called particle
states) outside the Fermi sea. In the present work the
$\rho_{MFA}({\bf r},{\bf r}')$ was constructed by using
single-particle wave functions that originated from the HO
potential.

\subsection{Natural orbitals representation of the OBDM}

A model-independent way to define a set of single-particle wave
functions  and occupation probabilities from the correlated  $\rho
({\bf r},{\bf r}')$ is to use the natural orbital representation
suggested, by L\"{o}wdin~\cite{Lowdin-55}, according to the
following expansion
\begin{equation}
\rho ({\bf r},{\bf r}') = \sum_{\alpha} n_{\alpha}
\psi_{\alpha}^*({\bf r}) \psi_{\alpha}({\bf r'}). \label{NO-1}
\end{equation}
The normalized eigenfunctions $\psi_{\alpha}({\bf r})$ of $\rho
({\bf r},{\bf r}')$ are called natural orbitals and they  form a
complete orthogonal set. The eigenvalues $n_{\alpha}$, where $0
\leq n_{\alpha} \leq 1$ are called natural occupation numbers and they define
the  occupation probability of the NO $\psi_{\alpha}({\bf r})$.
Usually there are  orbitals $\psi_{\alpha}({\bf r})$ for which the
occupation numbers $n_{\alpha}$ are significantly larger
than the others. These are called hole state orbitals and form a
new Fermi sea, while the rest are called particle state
orbitals~\cite{Stoitsov-93,Lewart-88}.

The  density distribution in natural orbitals representation is of
the form
\begin{equation}
\rho ({\bf r}) = \sum_{\alpha} n_{\alpha} |\psi_{\alpha}({\bf
r})|^2 \label{DD-1}. \end{equation}
The normalization is satisfied via the sum rule
\begin{equation}
\sum_{\alpha} n_{\alpha} =A\quad \mbox{or}\quad \sum_{\alpha}
p_{\alpha} = 1, \label{norm-1} \end{equation}
where $p_{\alpha}=n_{\alpha}/A$ is the occupation probability of
the state $\alpha$.

We apply the one-body density matrix, evaluated for the various
$sp$ and $sd$ shell nuclei~\cite{Moustakidis-00}. The various
terms of $\rho ({\bf r},{\bf r}')$ are  functionals of the harmonic
oscillator orbitals depending on the size parameter $b$ and the
correlation parameter $\beta$ (or $y=\beta
b^2$)~\cite{Moustakidis-00}.

In general, the diagonalization of $\rho ({\bf r},{\bf r}')$,
according to definition (\ref{NO-1}), is achieved by solving the
equation
\begin{equation}
\int \rho ({\bf r},{\bf r'}) \psi_{\alpha}({\bf r'}) d {\bf r'} =
p_{\alpha} \psi_{\alpha}({\bf r}).
 \label{diag-1}
\end{equation}

For nuclei with total angular momentum $J=0$, the OBDM has to be
diagonalized with the $\{ljm\}$ subspace of the complete space of
NO. For the case of the closed shell nuclei their angular part is
explicitly determined
\begin{equation}
\psi_{\alpha}({\bf r}) = \psi_{nlm}({\bf r}) = \psi_{nl}( r)
Y_l^m(\vartheta,\varphi).
 \label{diag-2}
\end{equation}
Substituting Eq. (\ref{diag-2}) into (\ref{diag-1}) and
integrating over the angular variables, the equation for the
radial part of the NO becomes
\begin{equation}
\frac{4\pi}{2l+1}\int_0^{\infty} \rho_l (r, r') \psi_{nl}(r') r'^2 d r' = p_{nl}
\psi_{nl}(r).
 \label{diag-3}
\end{equation}
The kernels $\rho_l (r, r')$ are the coefficients of the Legendre
expansion of the OBDM~\cite{Moustakidis-00}
\begin{equation}
\rho ({\bf r},{\bf r'})=\rho (r, r', \cos\omega_{rr'})\, =\,
 \sum_{l=0}^{\infty} \rho_l (r, r')
P_l(\cos\omega_{rr'}),
 \label{diag-4}
\end{equation}
where
\begin{equation}
\rho_l(r,r')= \frac{2l+1}{2} \int_{-1}^1 \rho (r, r',
\cos\omega_{rr'}) P_l(\cos\omega_{rr'}) d  (\cos\omega_{rr'}).
 \label{diag-5}
\end{equation}

From the analytical expression of the OBDM, $\rho ({\bf r},{\bf
r'})=\rho (r, r', \cos\omega_{rr'})$, the analytical expression of
the expansion coefficients $\rho_l(r,r')$ were
found and  they have
the form~\cite{Moustakidis-00}
\begin{equation}
\rho_l(r,r') \simeq N \big[ \rho_{1l}(r,r') - \sum_{i=1}^3
\rho_{22l}(r,r',g_i) \big],
 \label{diag-6}
\end{equation}
where the one-body contribution of the expansion coefficients and the three terms $\rho_{22l}(r,r',g_i)$ $(i=1,2,3)$,
which come from the two-body contribution of the cluster expansion, have the form
\begin{equation}
\rho_{1l}(r,r')=\frac{1}{\pi^{3/2}b^3} \big[ (4\eta_{1s}+\eta_{2s}
\Pi_0(r_b,r'_b)) \delta_{l0} +\eta_{1p} \Pi_1(r_b,r'_b) \delta_{l1}
+ \eta_{1d} \Pi_2(r_b,r'_b) \delta_{l2}\big] \exp [-(r_b^2 +
{r'}_{b}^{2})/2 ],
\label{rho-l-1}
\end{equation}
\begin{align}
\rho_{22l}(r,r',g_i)&=\exp [-\,
\frac{1+3y}{2(1+y)}\,{r'}_b^2 -
{r}_{b}^{2}/2 ] \nonumber \\
&\times \big[ A_{0l}(r_b,r'_b,g_i) \delta_{l0} +
A_{1l}(r_b,r'_b,g_i) \delta_{l1} +
A_{2l}(r_b,r'_b,g_i)(\delta_{l0} +2\delta_{l2})/3\big], \quad i=1,2
\label{rho-22-g1}
\end{align}
\begin{equation}
 \rho_{22l}(r,r',g_3)=\frac{2l+1}{2} \sum_{k=0}^{4} B_{kl}(r_b,r'_b,g_3) \big[
 f_{1kl}(r_b,r'_b) e^{c_1} + f_{1kl}(r_b,r'_b) e^{c_2} \big].
 \label{rho-22-g3}
 \end{equation}
In Eqs.~(\ref{rho-l-1}), (\ref{rho-22-g1}) and (\ref{rho-22-g3})
 $\Pi_i,\ A_{kl}, \ B_{kl}, \ f_{1kl},\ f_{2kl},\ c_1$ and $c_2$ are
 polynomials of $r_b=r/b$ and $r'_b=r'/b$. It is noted that $\rho_{1l}(r,r')=\rho_{22l}(r,r',g_1)=
\rho_{22l}(r,r',g_2)=0$ for $l>2$, while $\
\rho_{22l}(r,r',g_3)\ne 0\ \ \forall l$.

Using the analytical expressions of
$\rho_l(r,r')$ we solved the eigenvalue equation (\ref{diag-3})
and as a result  the dependence of the occupation probabilities
$p_{nl}$ and the NO $\psi_{nl}(r)$ on the parameters $b$ and $y$ can be studied.

\section{One body density matrix and momentum distribution in nuclear matter}
The model we used is based on the Jastrow ansatz for the
ground state wave function of nuclear matter
\begin{equation}
\Psi({\bf r}_1,{\bf r}_2,...,{\bf r}_N)= \prod_{1 \leq i \leq  j
\leq N} f(r_{ij}) \, \Phi_0({\bf r}_1,{\bf r}_2,...,{\bf r}_N),
\label{Jastrow-1}
\end{equation}
where $r_{ij}=|{\bf r}_i-{\bf r}_j|$, $\Phi_0$ is a Slater
determinant (here constructed by plane waves with appropriate
spin-isospin factors, filling the Fermi sea) and $f(r)$ is  a
state-independent two-body correlation function.

A cluster expansion for the one-body density matrix $\rho ({\bf
r},{\bf r'})$ of the nuclear matter, in the framework of the low
order approximation (LOA), has been derived by Gaudin, Gillespie and
Ripka \cite{Gaudin71,Flyn84} for the Jastrow trial function
(\ref{Jastrow-1}) and has the form
\begin{eqnarray}
\rho_{\rm LOA} ({\bf r},{\bf r'})&=&\rho_{0} ({\bf r},{\bf
r'})+\int \left[f(|{\bf r}-{\bf r}_2|)f(|{\bf r'}-{\bf
r}_2|)-1\right] \nonumber \\
&\times&\left[4\rho_{0} ({\bf r}_2,{\bf r}_2) \rho_{0} ({\bf r},{\bf
r'})-\rho_{0} ({\bf r},{\bf r}_2) \rho_{0} ({\bf
r}_2,{\bf r'}) \right]d{\bf r}_2 \nonumber \\
&-&\int \int \left[f^2(|{\bf r}_2-{\bf
r}_3|)-1\right]\rho_{0} ({\bf r},{\bf r}_2)\nonumber \\
&\times&\left[4\rho_{0} ({\bf r}_2,{\bf r'}) \rho_{0} ({\bf
r}_3,{\bf r}_3)-\rho_{0} ({\bf r}_2,{\bf r}_3) \rho_{0} ({\bf
r}_3,{\bf r'}) \right]d{\bf r}_2 d{\bf r}_3, \label{OBDM-LOA}
\end{eqnarray}
where $\rho_{0} ({\bf r},{\bf r'})$ is the density matrix
corresponding to the wave function $ \Phi_0$ (that is $f(r)=1$).
Formula (\ref{OBDM-LOA})  is also applicable to finite nuclei
\cite{Bohigas-80,Stoitsov-93}. In the case of uniform and infinite
nuclear matter $\rho_{\rm LOA} ({\bf r},{\bf r'})$ becomes only  a
function  of $u=|{\bf r}-{\bf r}'|$. In addition it is easy
to prove that \cite{Gaudin71,Flyn84}
\begin{equation}
\rho_{\rm LOA}(u)=\frac{4}{(2\pi)^3} \int^{{\bf k}_F} e^{-i {\bf
k}{\bf u}} d{\bf k}. \label{ru-1}
\end{equation}
Obviously $\rho_{\rm LOA}(0)=\rho_0$, where $\rho_0$ is the
density of the nuclear matter. The corresponding momentum
distribution is given by the Fourier transform
\begin{equation}
n_{\rm LOA}(k)=\frac{1}{4}\int  \rho_{\rm LOA}(u)e^{i {\bf k}{\bf
u}}d{\bf u}. \label{nk-loa-1}
\end{equation}

Adopting the gaussian model for the correlation function $f(r)$
\begin{equation}
f(r)=1-\exp[-\beta r^2], \label{beta-Gaus}
\end{equation}
the momentum distribution can be written
\cite{Flyn84}
\begin{equation}
n_{\rm LOA}(k)=\theta(k_F-k) \left[ 1-k_{\rm dir}+Y(k,8) \right]+
8 \left[ k_{\rm dir}Y(k,2)-[Y(k,4)]^2 \right], \label{mn-mom-1}
\end{equation}
where
\begin{equation}
c_{\mu}^{-1}Y(k,\mu)=
\frac{e^{-\tilde{k}_{+}^{2}}-e^{-\tilde{k}_{-}^{2}}}{2\tilde{k}}
+\int_{0}^{\tilde{k}_{+}} e^{-y^2}  dy + {\rm sgn}(\tilde{k}_{-})
\int_{0}^{\mid \tilde{k}_{-} \mid} e^{-y^2} dy,
\end{equation}
and
\begin{equation}
    c_{\mu}=\frac{1}{8\sqrt{\pi}}\left(\frac{\mu}{2}\right)^{3/2},
    \quad \tilde{k}=\frac{k}{\sqrt{\beta\mu}} , \quad
    \tilde{k}_{\pm}=\frac{k_{F}\pm k}{ \sqrt{\beta \mu}}, \quad
    \mu=2,4,8,
\end{equation}
while ${\rm sgn}(x)=x/|x|$.

The dimensionless Jastrow wound
parameter $k_{\rm dir}$, which  can serve as a rough measure of
correlations and the rate of convergence of the cluster expansion,
is defined as
\begin{equation}
    k_{\rm dir}=\rho_0 \int [f(r)-1]^2 \, d{\bf r}, \label{eq-kdir}
\end{equation}
where $\rho_0$ is the density of nuclear matter,
which is related to the Fermi wave number $k_F$ by the  equation $\rho_0=2k_F^3/(3\pi^2)$.
The normalization condition for the momentum distribution is
\begin{equation}
\int_{0}^{\infty} n_{\rm LOA}(k)k^2 dk=\frac{1}{3}\, k_{F}^{3}.
\end{equation}

From  Eq.~(\ref{eq-kdir}) we obtain the following relation between
the wound parameter $k_{\rm dir}$ and the correlation parameter
$\beta$
\begin{equation}
k_{\rm dir}=\frac{1}{3 \sqrt{2\pi}}\left(\frac{k_F}{\sqrt{\beta}}
\right)^3.
\end{equation}

It is clear that large values of  $k_{\rm dir}$ imply strong
correlations and poor convergence of the  cluster expansion.
A reasonable interval of the values of the correlation parameter
$\beta$ is: $1.13 \ {\rm fm}^{-2} \leq \beta \leq  7 \ {\rm fm}^{-2} $ \cite{Flyn84}.
That range corresponds to $ 0.3 \geq k_{\rm dir} \geq 0.02 $. In our calculations we considered
the value $\rho_0=0.182$ fm$^{-3}$ for the density of uniform,
spin-isospin-saturated nuclear matter which corresponds to
$k_F=1.3915$ fm$^{-1}$. This value is close to $k_F$ of
symmetrical nuclear matter at equilibrium density. It is mentioned that Eq.~(\ref{OBDM-LOA}) and (\ref{mn-mom-1})
which are applicable to nuclear matter at lower effective
densities, can also be applied reliably to nuclei \cite{Gaudin71,Flyn84}.
Thus, a comparison of the correlated parameter i.e. of the effect of SRC in finite nuclei and in nuclear matter
can be made by considering a value of the nuclear matter density close to equilibrium density
of symmetrical nuclear matter.

\section{Results and Discussion}

We used the one-body density matrix, calculated in our previous
work~\cite{Moustakidis-00}, in order to construct a corresponding
set of natural  orbitals and natural occupation numbers for
various $sp$ and $sd$ shell nuclei, with $N=Z$. In that model
there are two free parameters, the size parameter $b$ and the
correlation parameter $\beta$ (or $y=\beta b^2$). These have been
determined by least squares fit to the experimental form factor
and are given in Table~I of Ref.~\cite{Moustakidis-00}. It is worth
noting  that both the NO and the NON depend strongly on the
correlation parameter $y$ and one may study extensively the effect
of the SRC on the above quantities. However the purpose of the
present work is to use fixed parameters from the previous work and to
compare NO and NON with other similar studies and the existing
experimental data.

First we define the occupation ratio $\eta_a$ of the state
$\alpha$ as
\begin{equation}
\eta_a= \frac{ \rm number \ of \ nucleons \ occupy \ the \ state \
\alpha }{\rm maximum \ capacity \ of \ the \ state\  \alpha }.
\label{}
\end{equation}
The ratio $\eta_{\alpha}$ is related to the occupation probability
$p_{\alpha}$ according to

\begin{equation}
\eta_{\alpha}=\frac{p_{\alpha} \cdot A}{\rm maximum \ capacity \
of \ the \ state \ \alpha}.\label{}
\end{equation}
In view of the above definition the occupation ratios of the
various NO states will be given by the expressions
\begin{equation}
\eta_{s}=\frac{p_{s}\cdot A}{4}, \quad \eta_{p}=\frac{p_{p}\cdot
A}{12}, \quad  \eta_{d}=\frac{p_{d}\cdot A}{20} \quad
\eta_{f}=\frac{p_{f}\cdot A}{28}, \quad \eta_{g}=\frac{p_{g}\cdot
A}{36}. \label{}
\end{equation}

The values of the occupation ratio $\eta_{\alpha}$ of various
states for various $sp$ and $sd$ shell nuclei are
presented in Table~\ref{tb1}. It is noted that in the MFA (absence of
correlations) the corresponding value $\eta_{\alpha}^{\rm{MFA}}$ for the
closed shell nuclei is $1$ for the hole states and $0$ for the
particle states. For non closed shell nuclei, we considered
that for the MFA hole states  $1p$ for $^{12}$C and  $1d$ for
$^{24}$Mg, $^{28}$Si, $^{32}$S and $^{36}$Ar the occupation ratios $\eta_{\alpha}^{\rm{MFA}}$
are: $0.6667$, $0.4$, $0.6$, $0.8$ and $1$, respectively.

The values of $\eta_{\alpha}$, given in Table~\ref{tb1},  are compared with experimental
data~\cite{Kramer-90} and the theoretical estimate obtained for
$^{40}$Ca in Ref.~\cite{Stoitsov-93}. It should be noted that  the values of the occupation
ratios $\eta_{\alpha}$ are sensitive on the various elements of
the correlated OBDM as well as on the strength  of the
correlations. The use of different methods in the evaluation of
the OBDM may lead to quantitatively different values of
$\eta_{\alpha}$.

In a few cases arose non-physical negative values  for some
occupation ratios $\eta_a$ while in two cases (1d and 2s states of
$^{40}$Ca) the occupation ratios $\eta_{1d}$ and $\eta_{2s}$
surpass slightly the upper limit, that is $\eta_{1d},
\eta_{2s}>1$. Those results originated from the breaking of the
$A$-representability \cite{Kryachko-90}  of the OBDM applied in
the present work. The breaking of the $A$-representability depends
both on the form of the correlated one-body density matrix as well
as on the strength of the correlations which were introduced via
the correlation parameter $y$. In general the breaking of the $A$-
representability expresses, in a quantitative as well as
qualitative way,  the pathology of the approximations made in the
expansion of the density matrix $\rho({\bf r},{\bf r}')$. One may
be able to restore this problem by excluding all states with
negative $\eta_a$ and eventually renormalize the resulting
A-representable density matrix $\rho({\bf r},{\bf r}')$
\cite{Stoitsov-93} but this procedure is out of the purpose of the
present work.

The depletion of the hole states, which is defined as

\begin{equation}
D_a=\left( 1-\frac{\eta_{\alpha}}{\eta_{\alpha}^{\rm{MFA}}} \right)\cdot 100\,(\%)  \, ,
\label{Da-nuclei}
\end{equation}
and the Fermi sea depletion $D_F$, which is defined as
\begin{equation}
D_F=\frac{1}{A}
\left(4\eta_{1s}^{\rm{MFA}}D_{1s}+12\eta_{1p}^{\rm{MFA}}D_{1p}
+20\eta_{1d}^{\rm{MFA}}D_{1d}+4\eta_{2s}^{\rm{MFA}}D_{2s}\right) \,(\%) \, ,
\label{Total-nuclei}
\end{equation}
are given in Table~\ref{tb2}.

It is instructive to compare the present values of the  FSD
with those originating from other theoretical and experimental
studies. Actually, the  FSD, compared with the specific
depletion of the various states, can be measured, at least
indirectly or with model dependent analysis, in relative experiments.
The value of the  nuclear FSD, in the present work,
according to Table~\ref{tb2} is: $5.16 \%$ for $^4$He, $3.41 \%$ for
$^{12}$C, $4.00 \%$ for $^{16}$O, $1.59 \%$ for $^{24}$Mg, $2.85
\%$ for $^{28}$Si, $2.52 \%$ for $^{32}$S, $5.38 \%$ for $^{36}$Ar
and $5.91 \%$ for $^{40}$Ca. It is obvious that the value of the FSD is higher in the
closed shell nuclei $^4$He, $^{16}$O and
$^{40}$Ca where the SRC are stronger compared to the
neighbourly open shell nuclei. The above statement can only be
confirmed by future experimental studies of the Fermi sea
depletion which will include both closed and open shell nuclei.
The present results concerning the FSD of closed shell nuclei, are
comparable to those evaluated by Stoitsov {\it et
al.}~\cite{Stoitsov-93}. In the mentioned reference the depletions
of the Fermi sea were $5.7 \%$, $3.9 \%$, and $6.3 \%$ for $^4$He,
$^{16}$O and $^{40}$Ca, respectively. However, for the case of
$^{40}$Ca, the depletion is smaller compared to the value $9.4\%$
obtained in the analysis of the experimental data from $(e,e'p)$
and ($d,{^3}$He) reactions on the $^{40}$Ca nucleus
\cite{Kramer-90} .

\begin{figure}[hbtp]

  \centerline{\hbox{ \hspace{0.0in}
    \epsfxsize=2.75in
    \epsffile{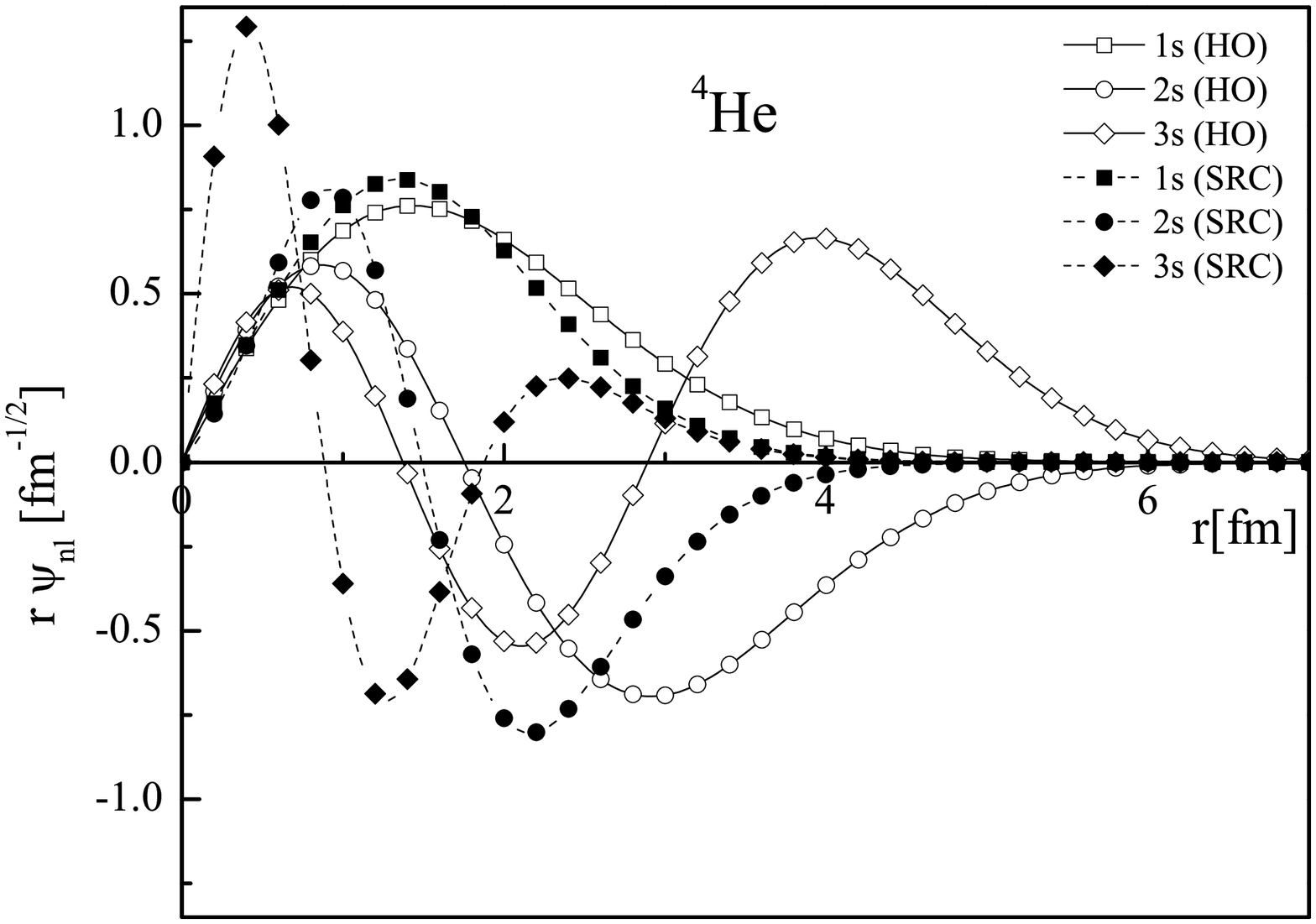}
    \hspace{-0.35in}
    \epsfxsize=2.75in
    \epsffile{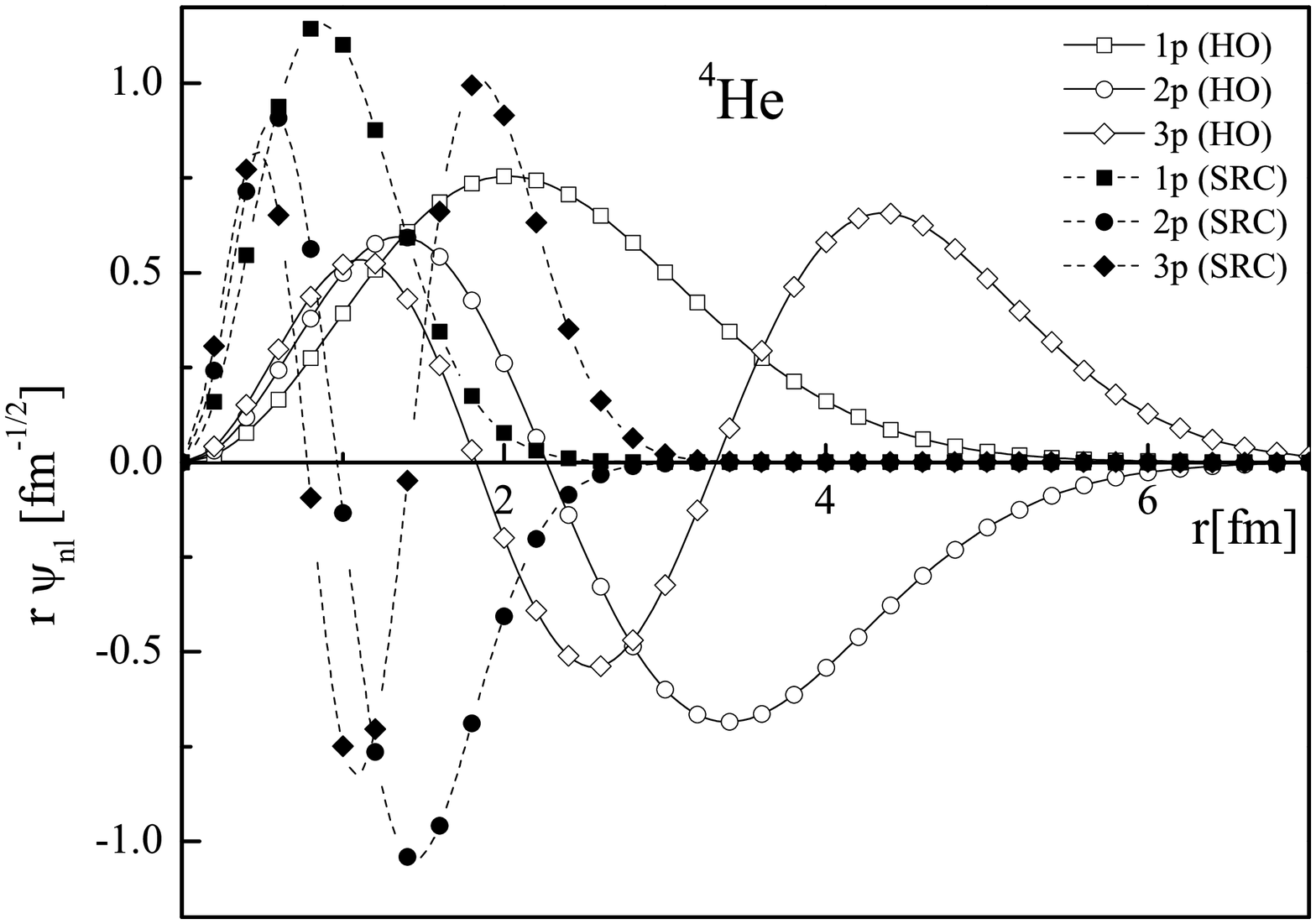}
    \hspace{-0.35in}
    \epsfxsize=2.75in
    \epsffile{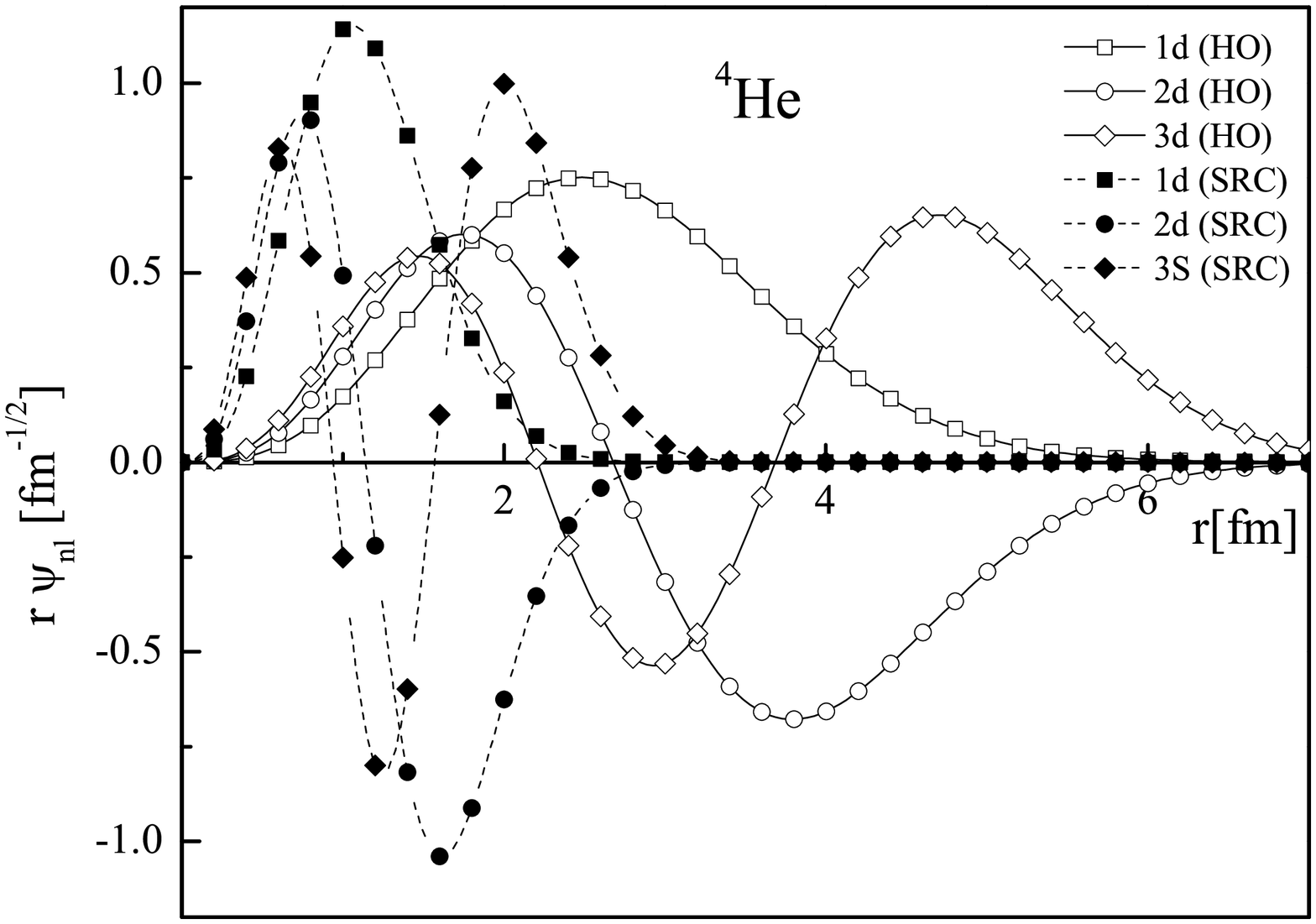}
    }
  }
    \vspace{-10px}
  \centerline{\hbox{ \hspace{0.0in}
    \epsfxsize=2.75in
    \epsffile{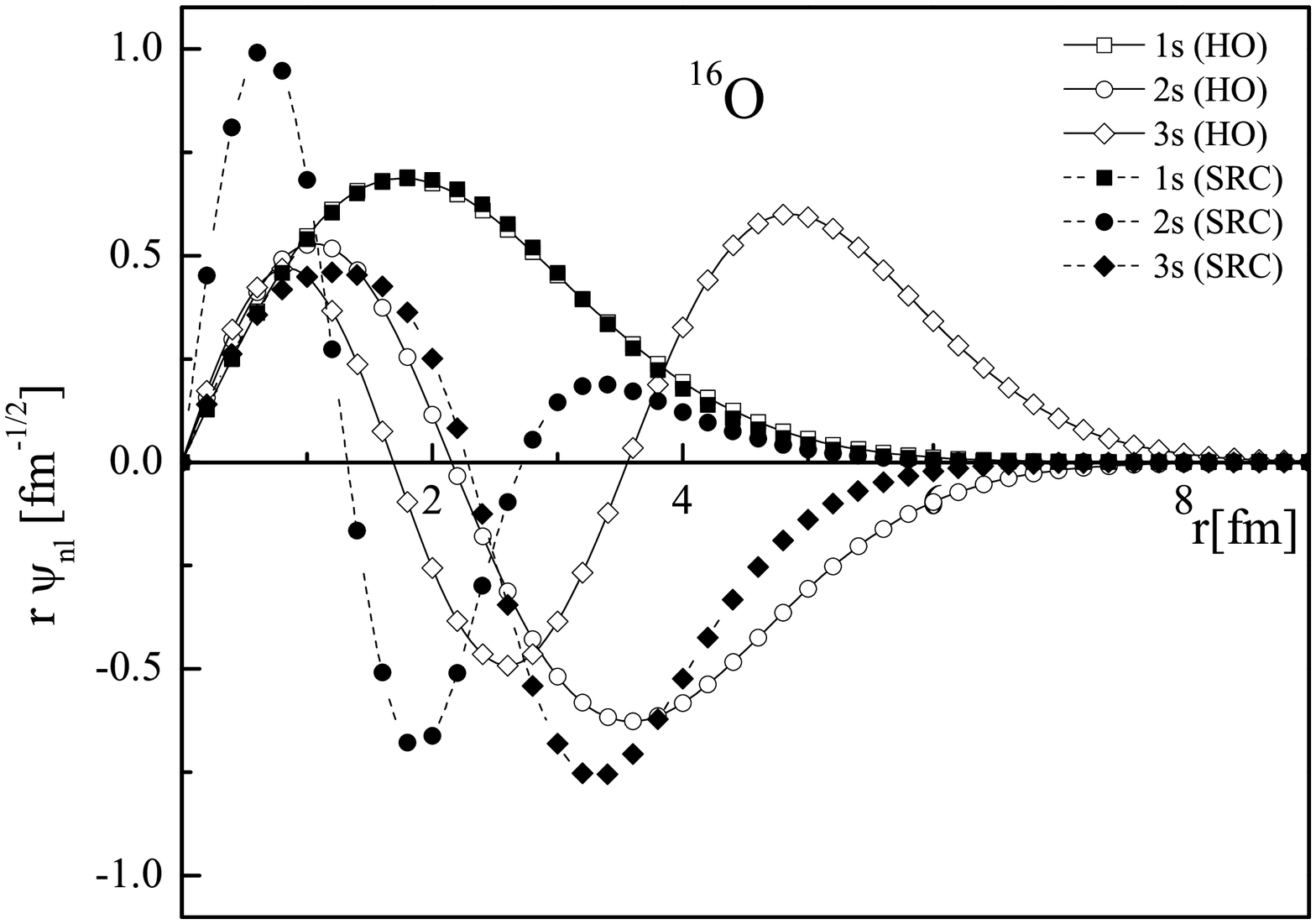}
    \hspace{-0.35in}
    \epsfxsize=2.75in
    \epsffile{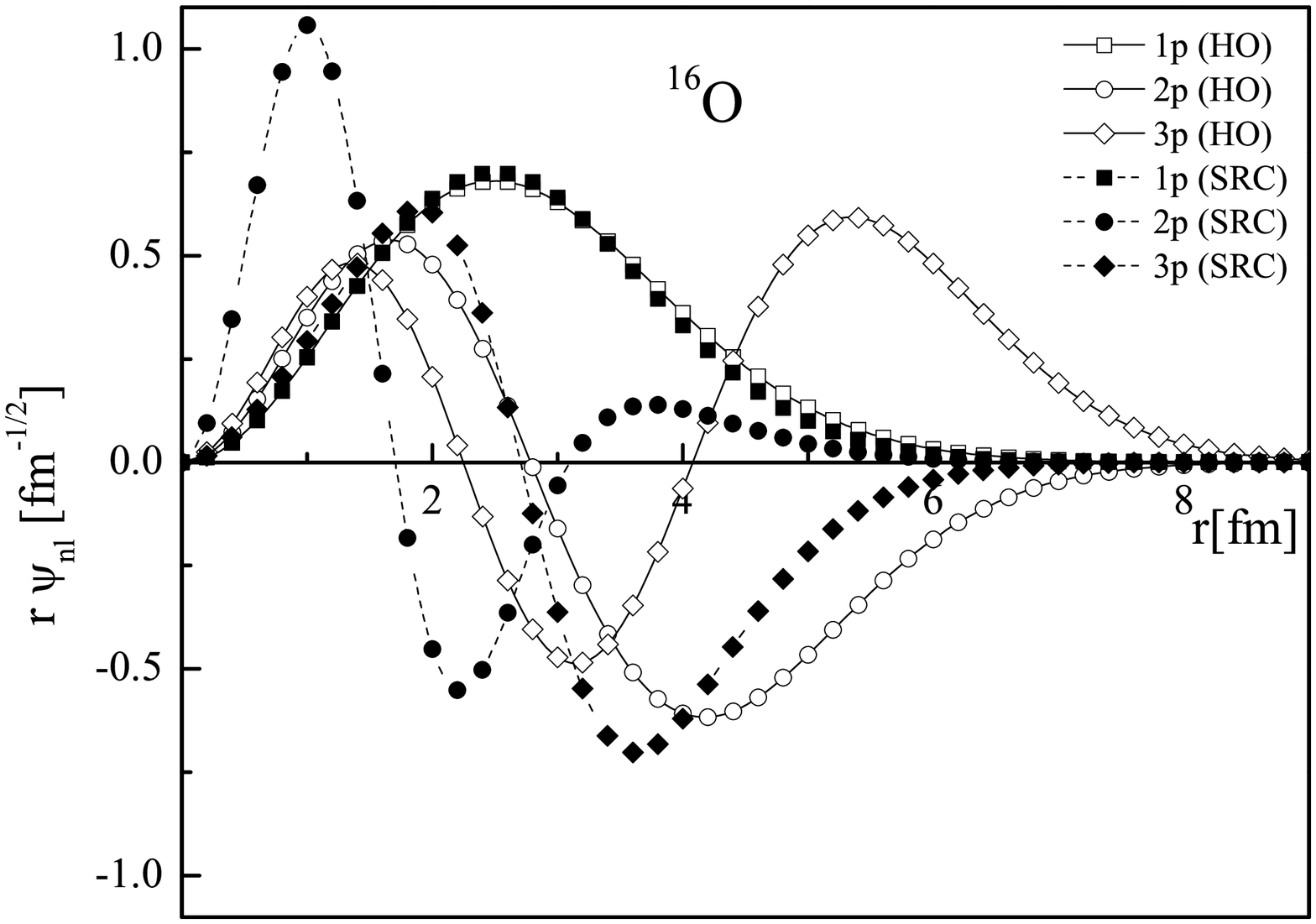}
    \hspace{-0.35in}
    \epsfxsize=2.75in
    \epsffile{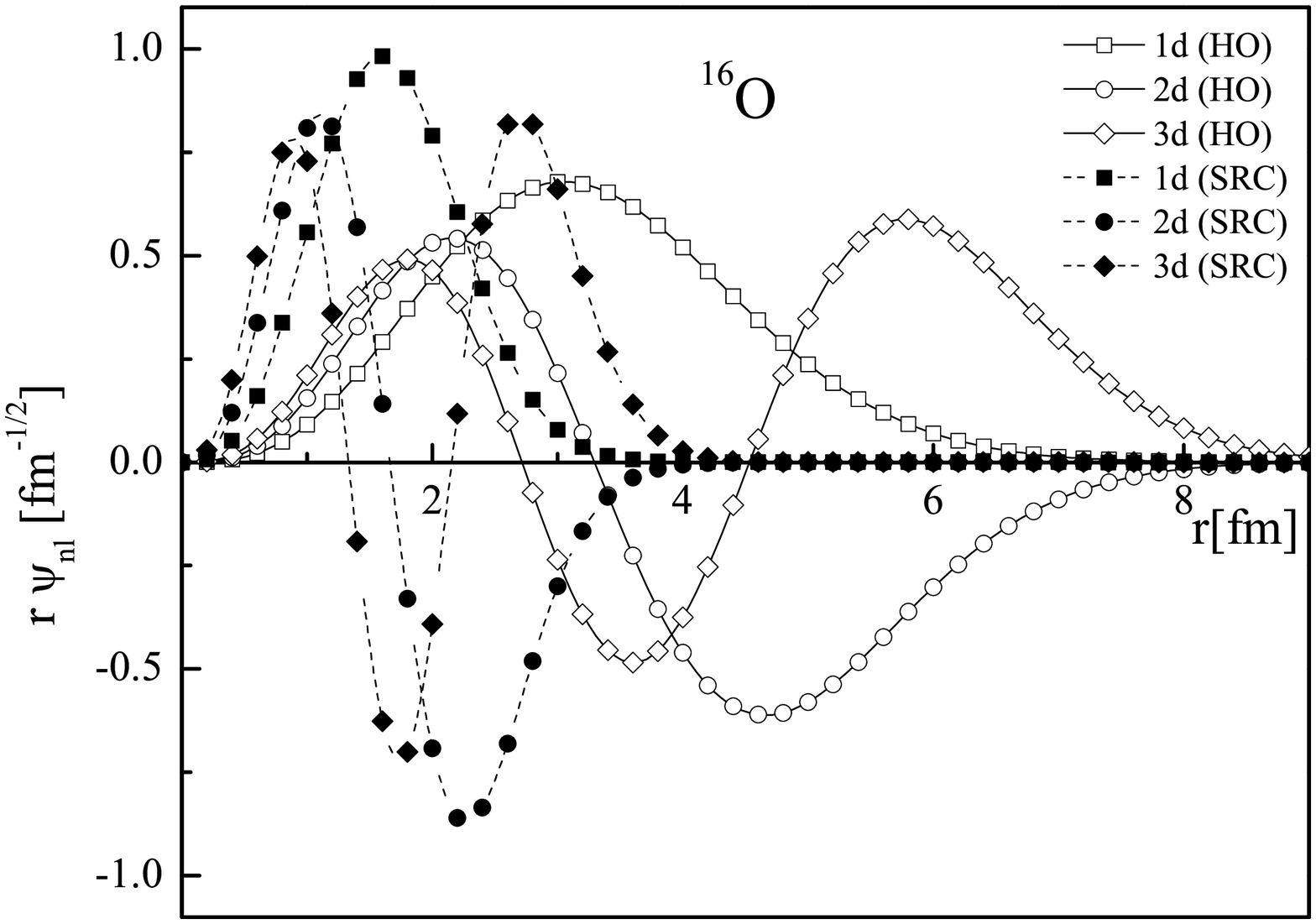}
    }
  }
    \vspace{-10px}
 \centerline{\hbox{ \hspace{0.0in}
    \epsfxsize=2.75in
    \epsffile{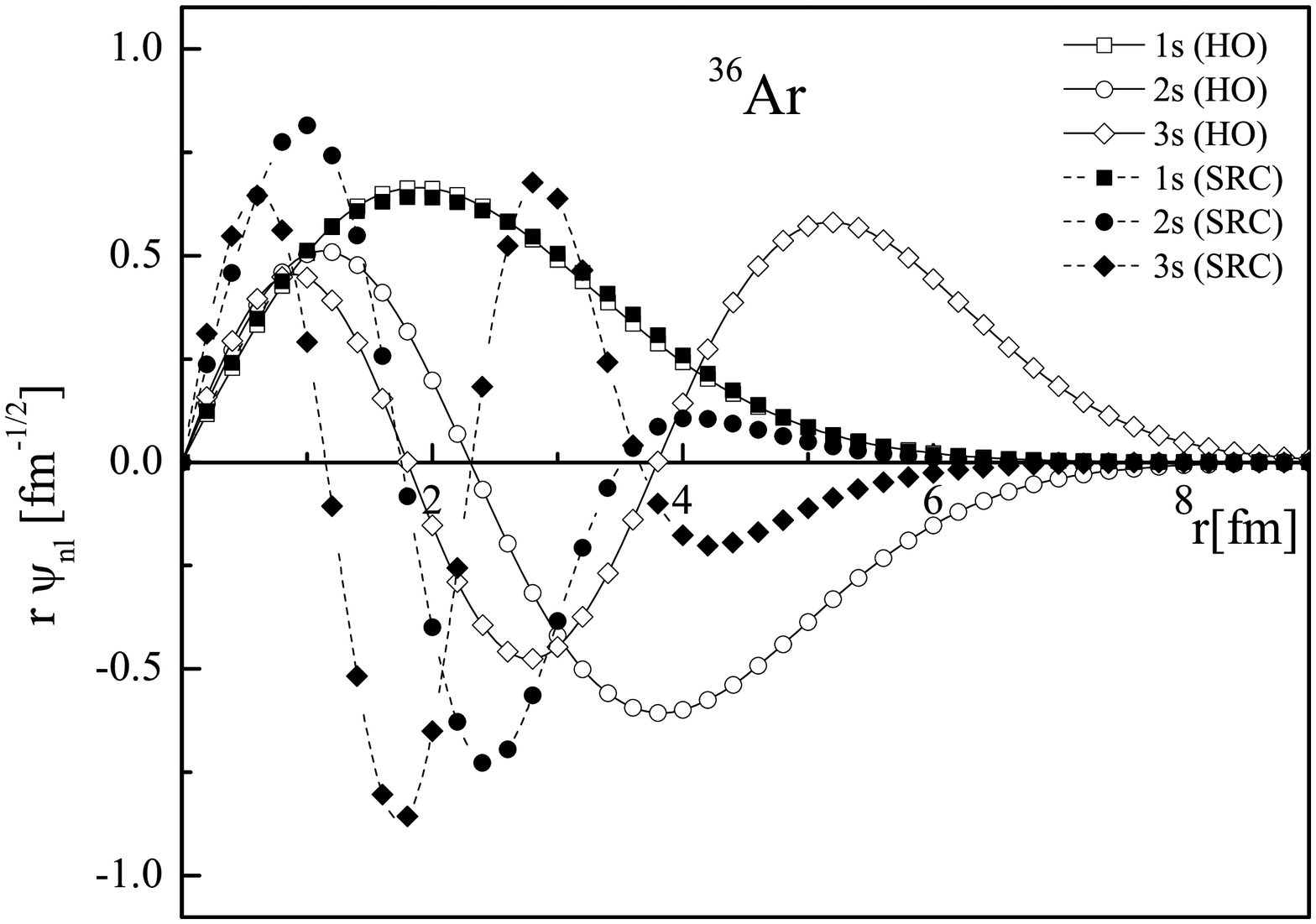}
    \hspace{-0.35in}
    \epsfxsize=2.75in
    \epsffile{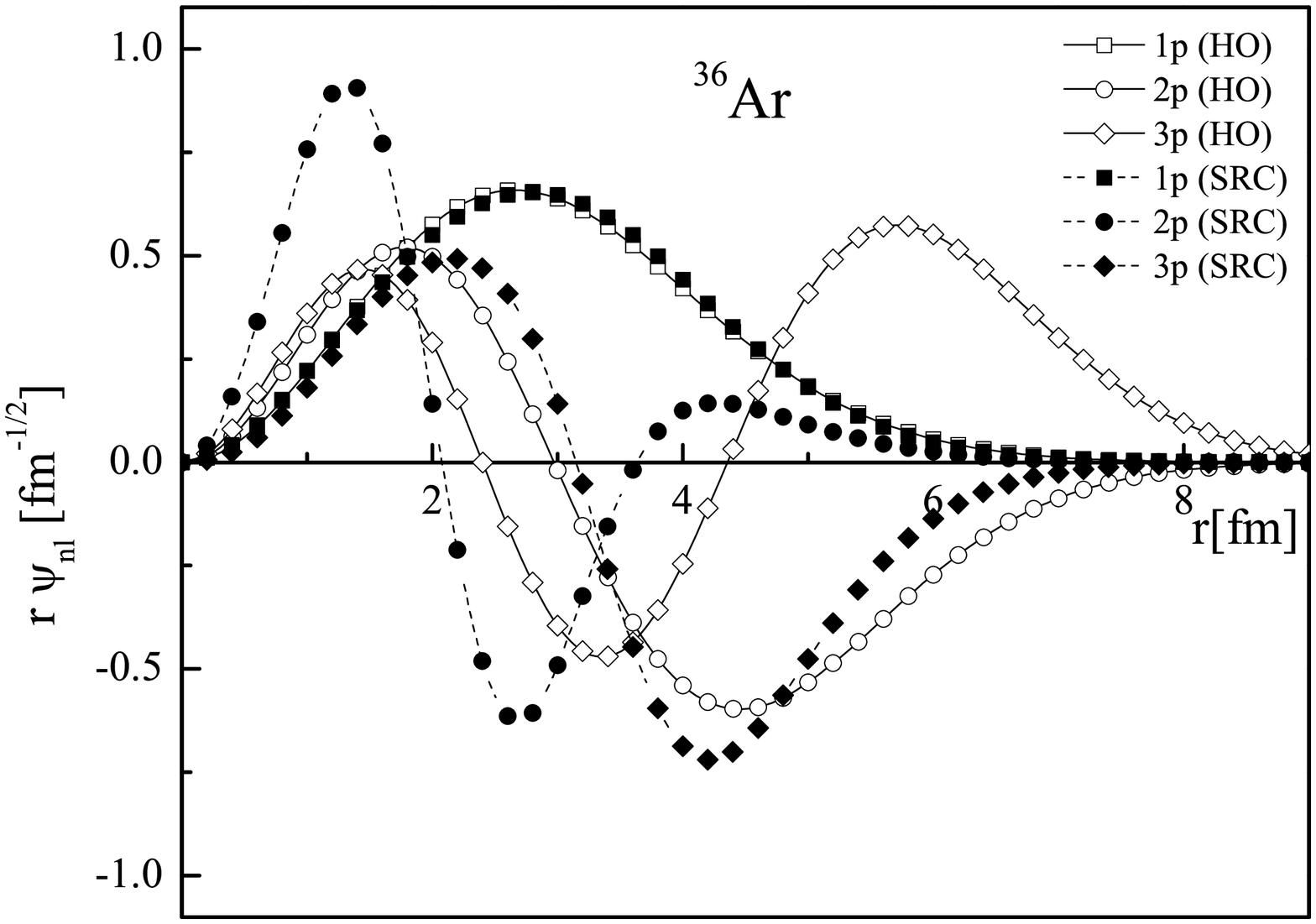}
    \hspace{-0.35in}
    \epsfxsize=2.75in
    \epsffile{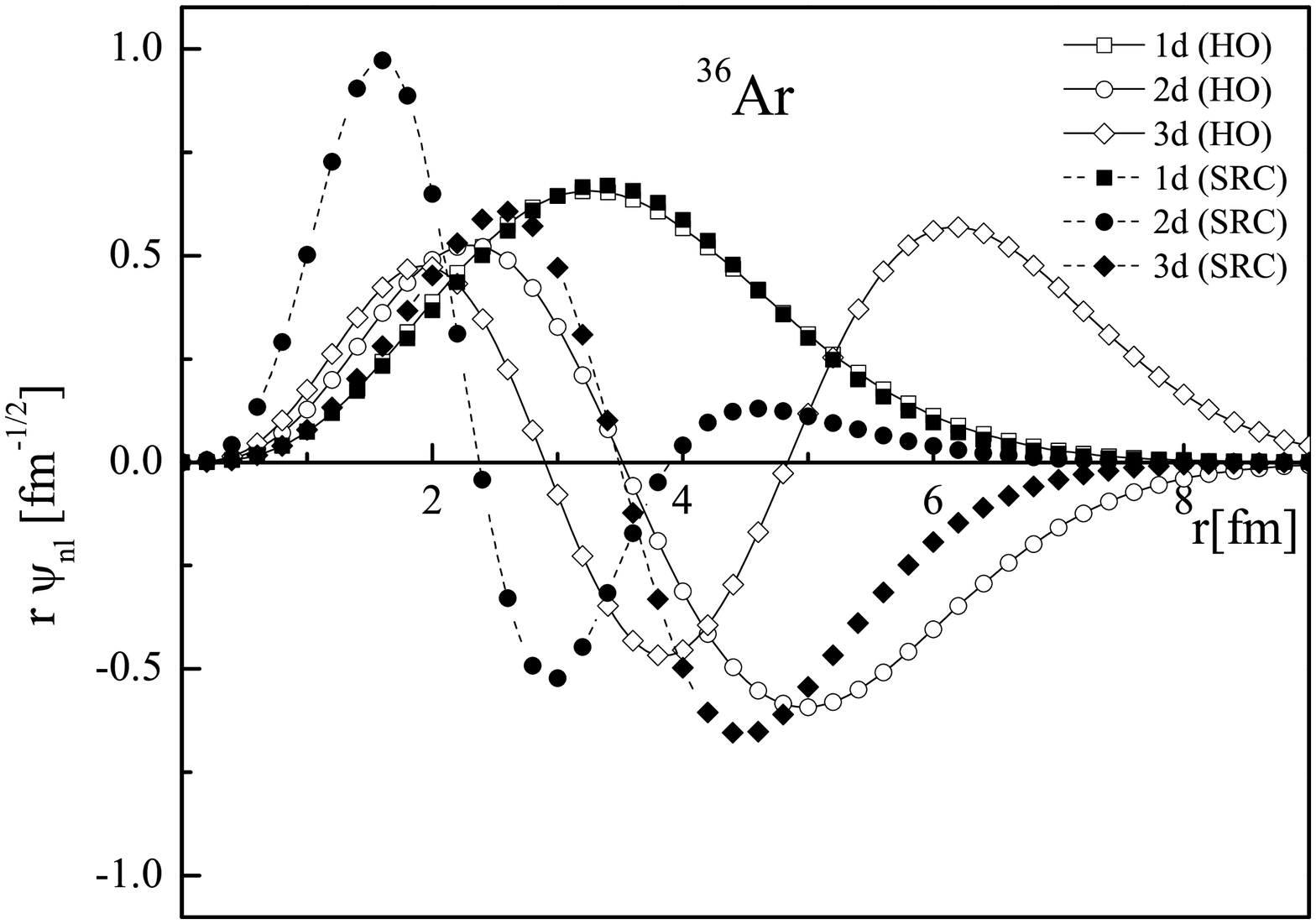}
    }
  }
    \vspace{-10px}
 \centerline{\hbox{ \hspace{0.0in}
    \epsfxsize=2.75in
    \epsffile{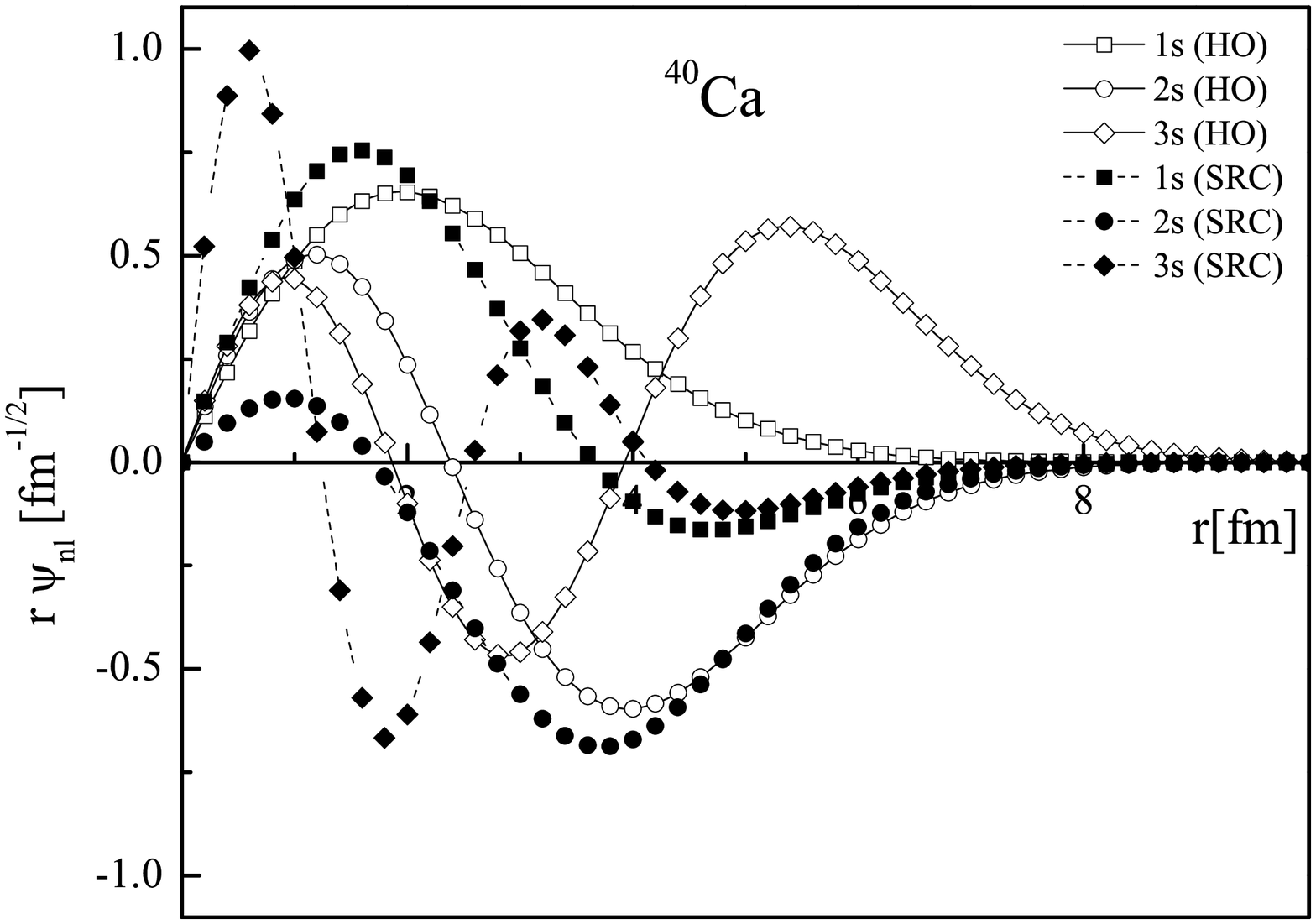}
    \hspace{-0.35in}
    \epsfxsize=2.75in
    \epsffile{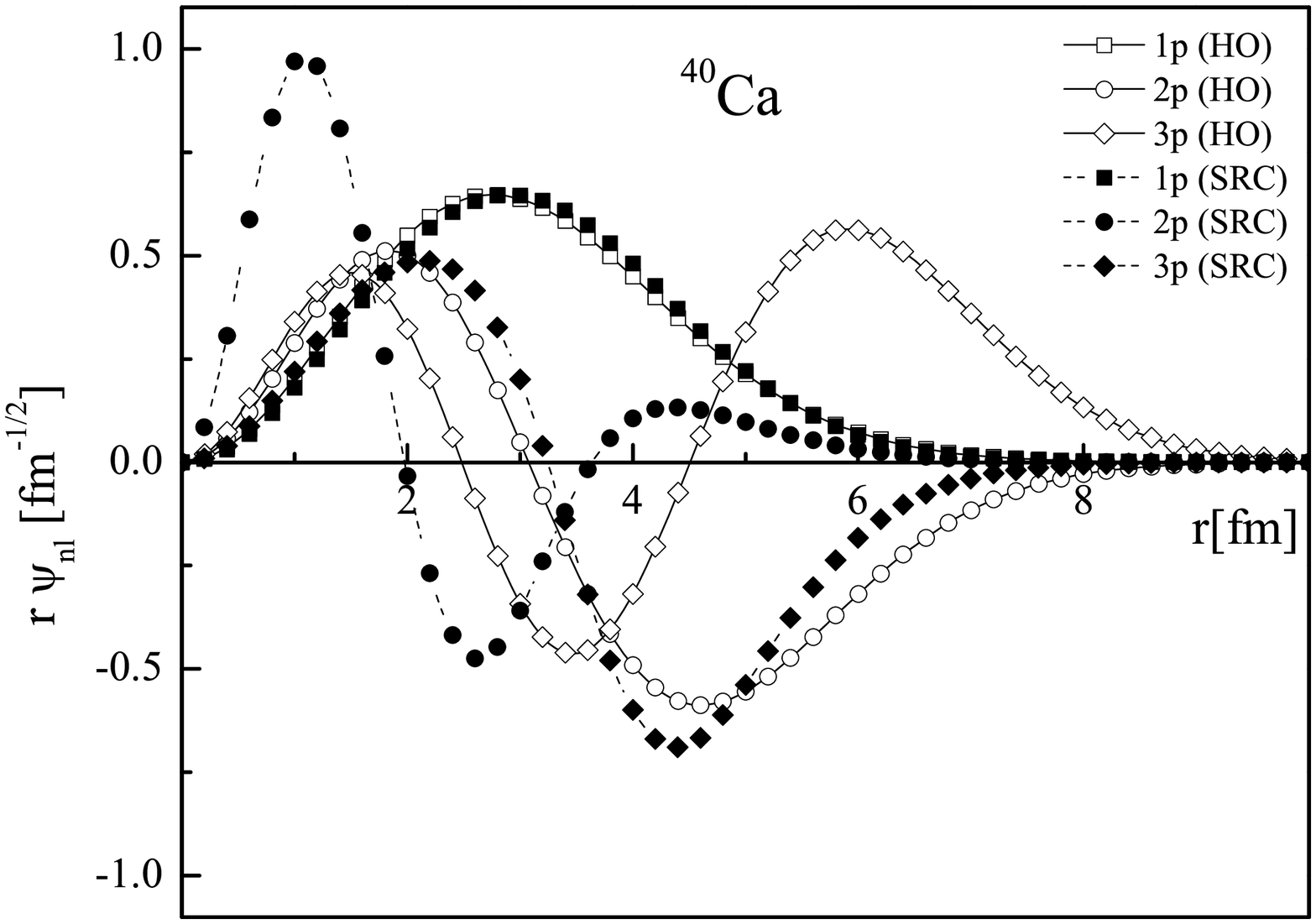}
    \hspace{-0.35in}
    \epsfxsize=2.75in
    \epsffile{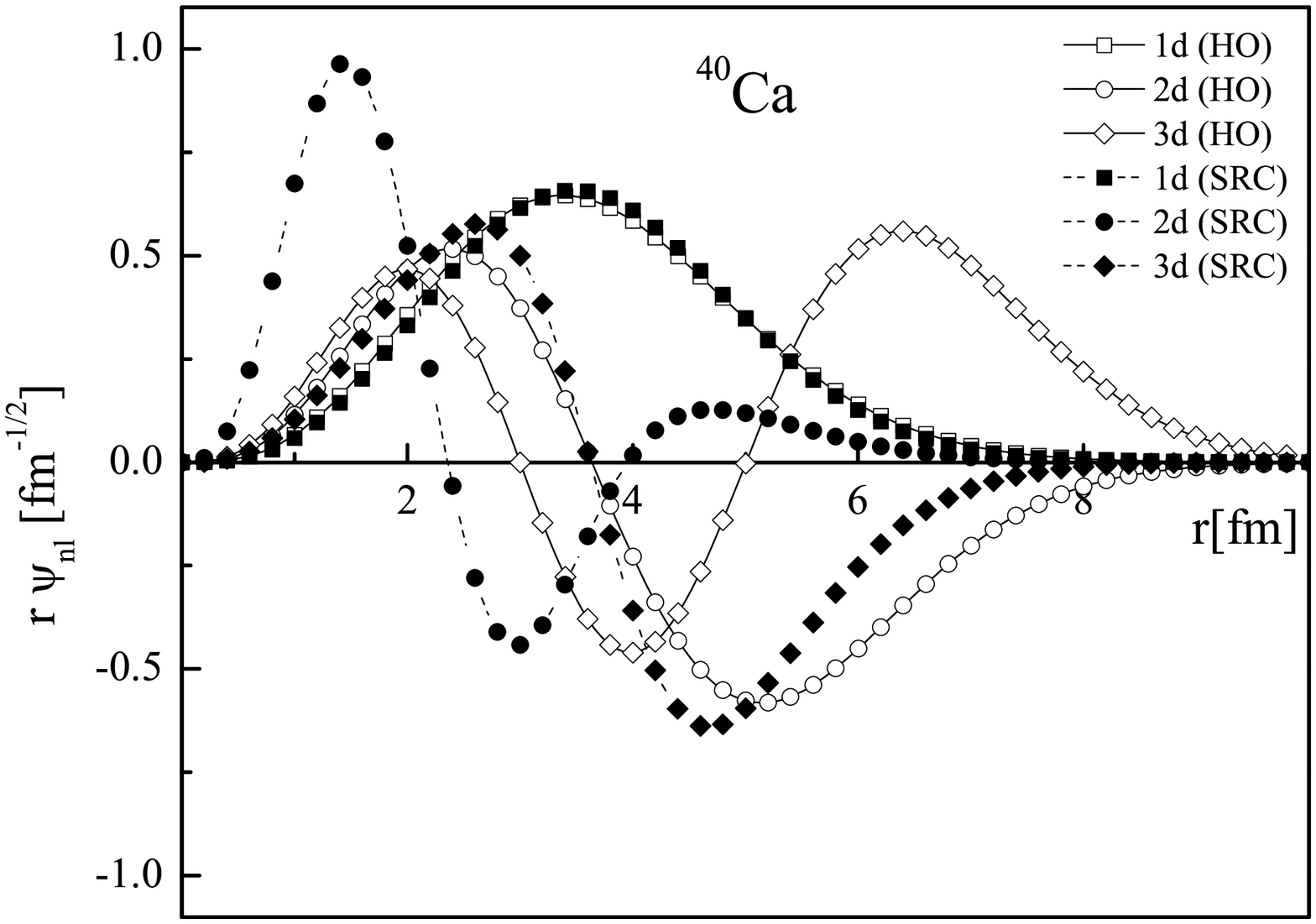}
    }
  }
\vspace{-1cm}
\caption{The natural orbitals of the particle- and hole-states (solid lines)
compared to the corresponding HO orbitals (dashed lines) for the
closed shell nuclei  $^4$He, $^{16}$O, $^{40}$Ca as well as for $^{36}$Ar.}
\label{fg1}

\end{figure}

\begin{figure}[hbtp]

  \centerline{\hbox{ \hspace{0.0in}
    \epsfxsize=2.75in
    \epsffile{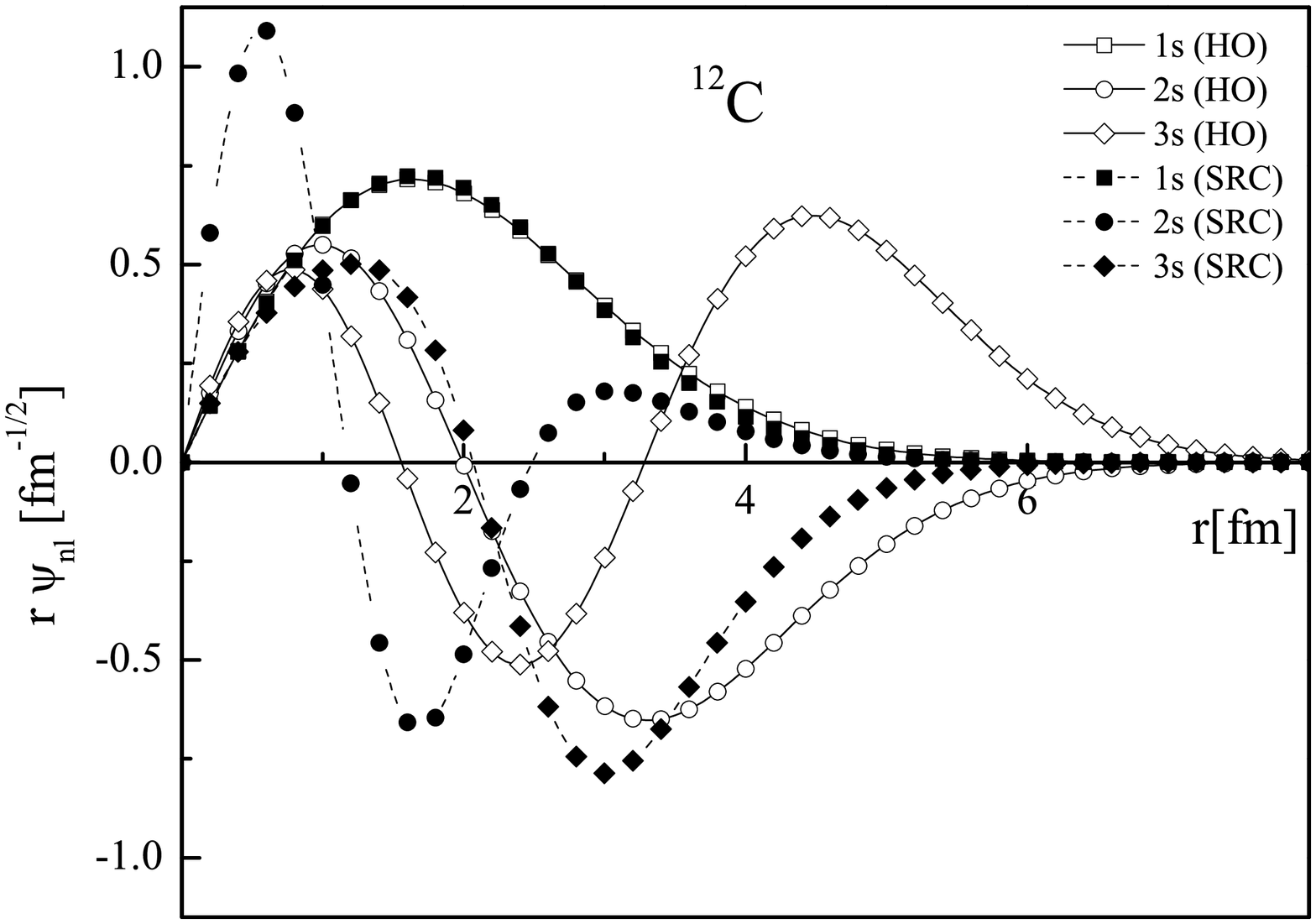}
    \hspace{-0.35in}
    \epsfxsize=2.75in
    \epsffile{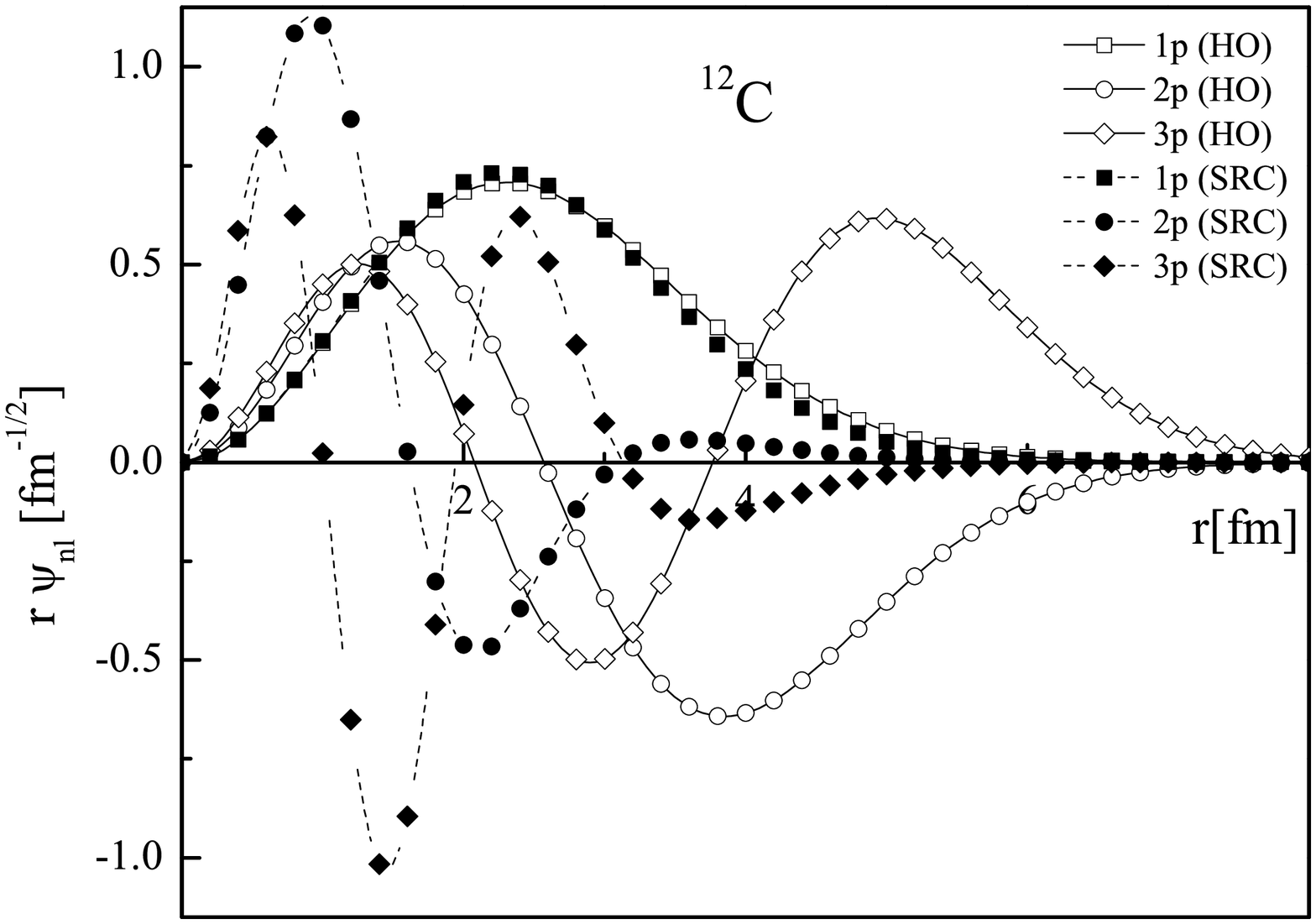}
    \hspace{-0.35in}
    \epsfxsize=2.75in
    \epsffile{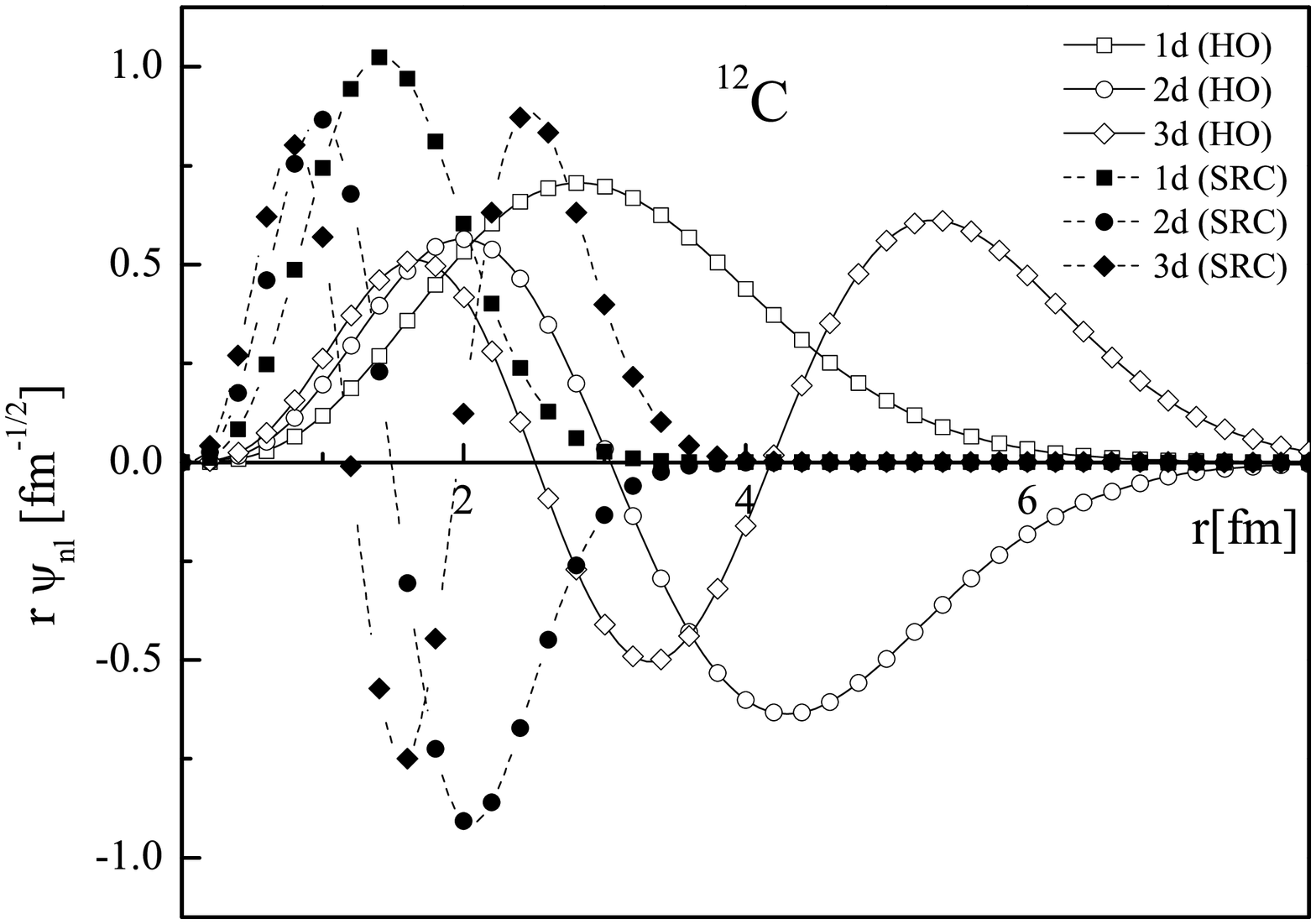}
    }
  }
    \vspace{-10px}
  \centerline{\hbox{ \hspace{0.0in}
    \epsfxsize=2.75in
    \epsffile{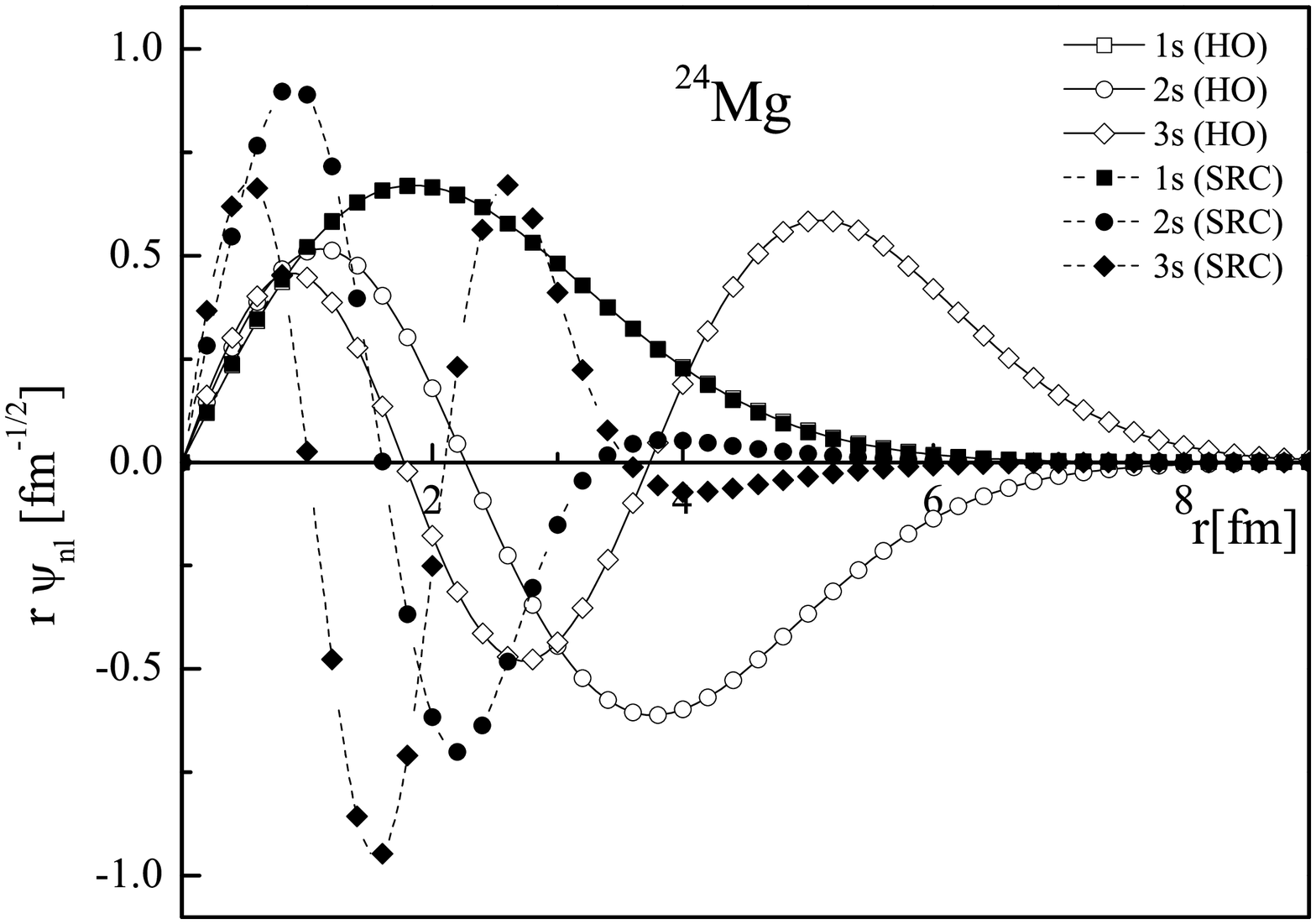}
    \hspace{-0.35in}
    \epsfxsize=2.75in
    \epsffile{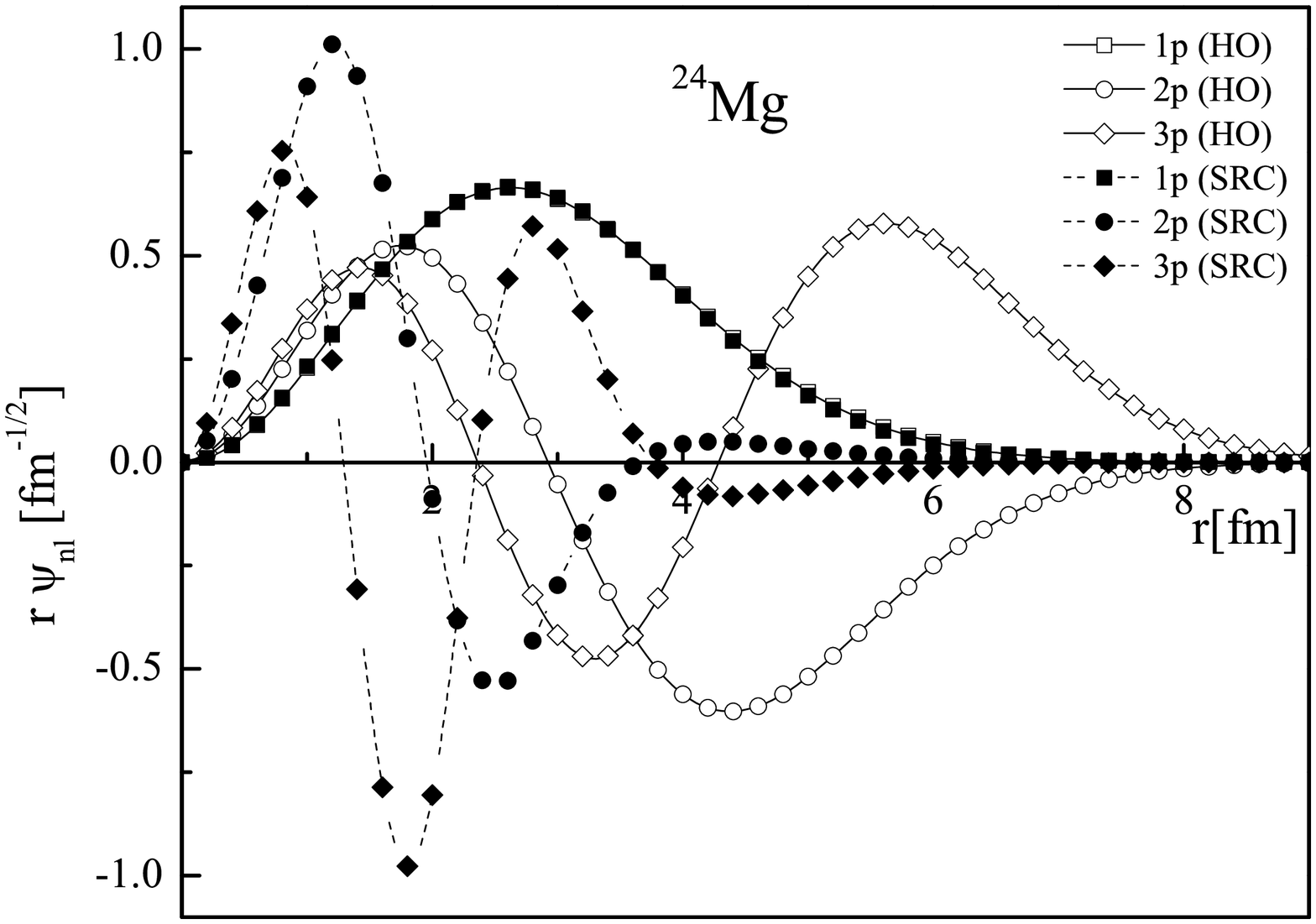}
    \hspace{-0.35in}
    \epsfxsize=2.75in
    \epsffile{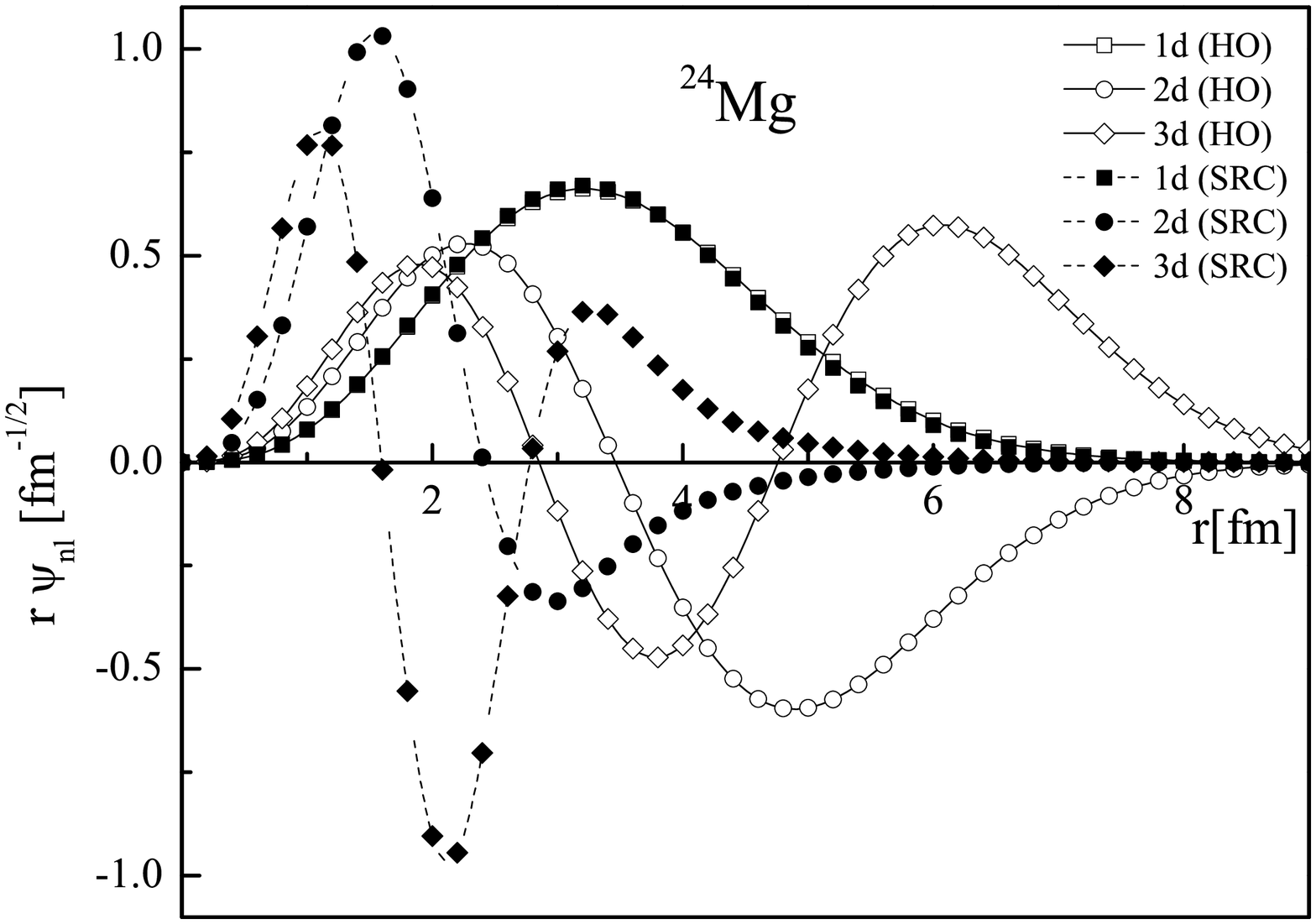}
    }
  }
    \vspace{-10px}
 \centerline{\hbox{ \hspace{0.0in}
    \epsfxsize=2.75in
    \epsffile{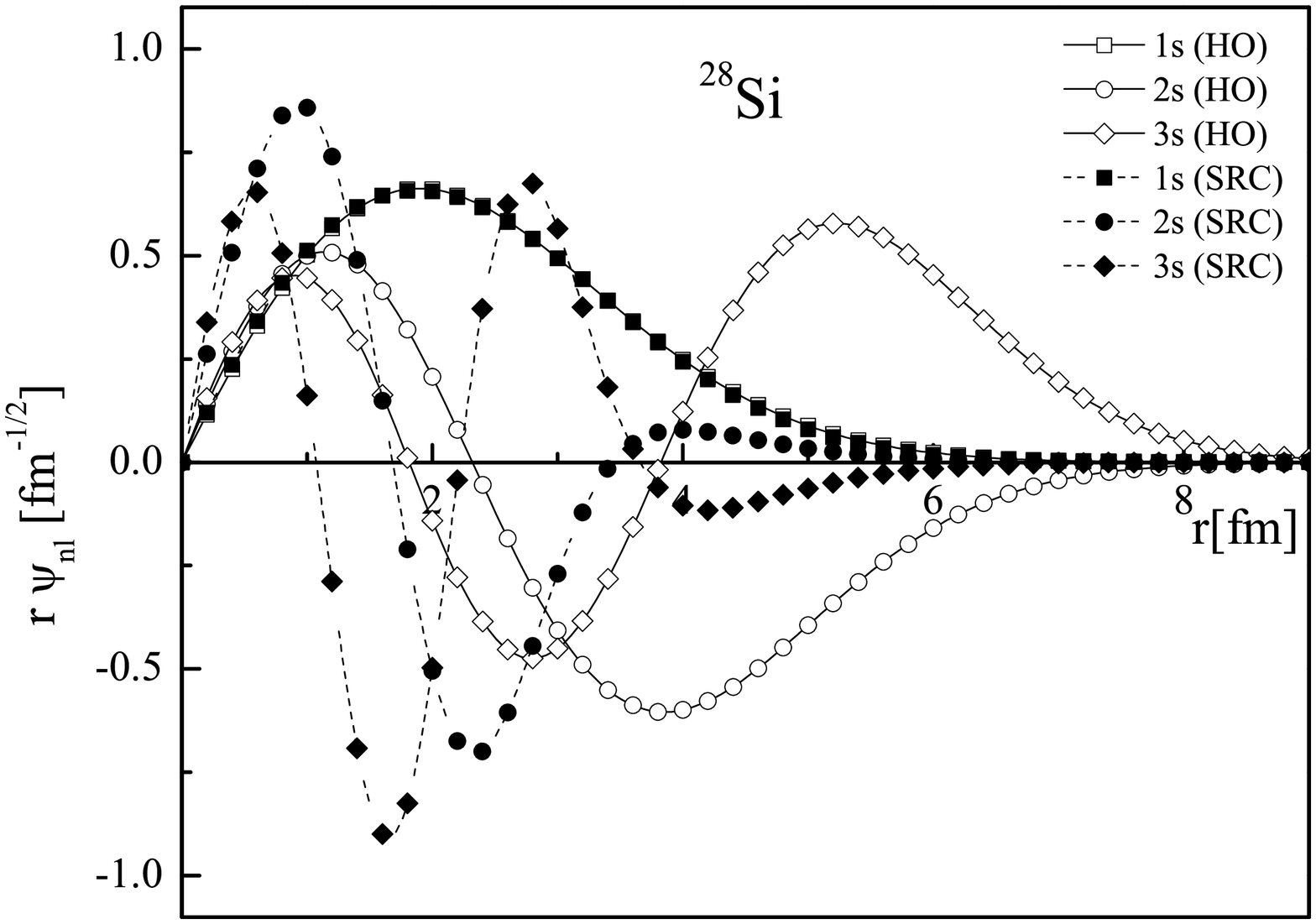}
    \hspace{-0.35in}
    \epsfxsize=2.75in
    \epsffile{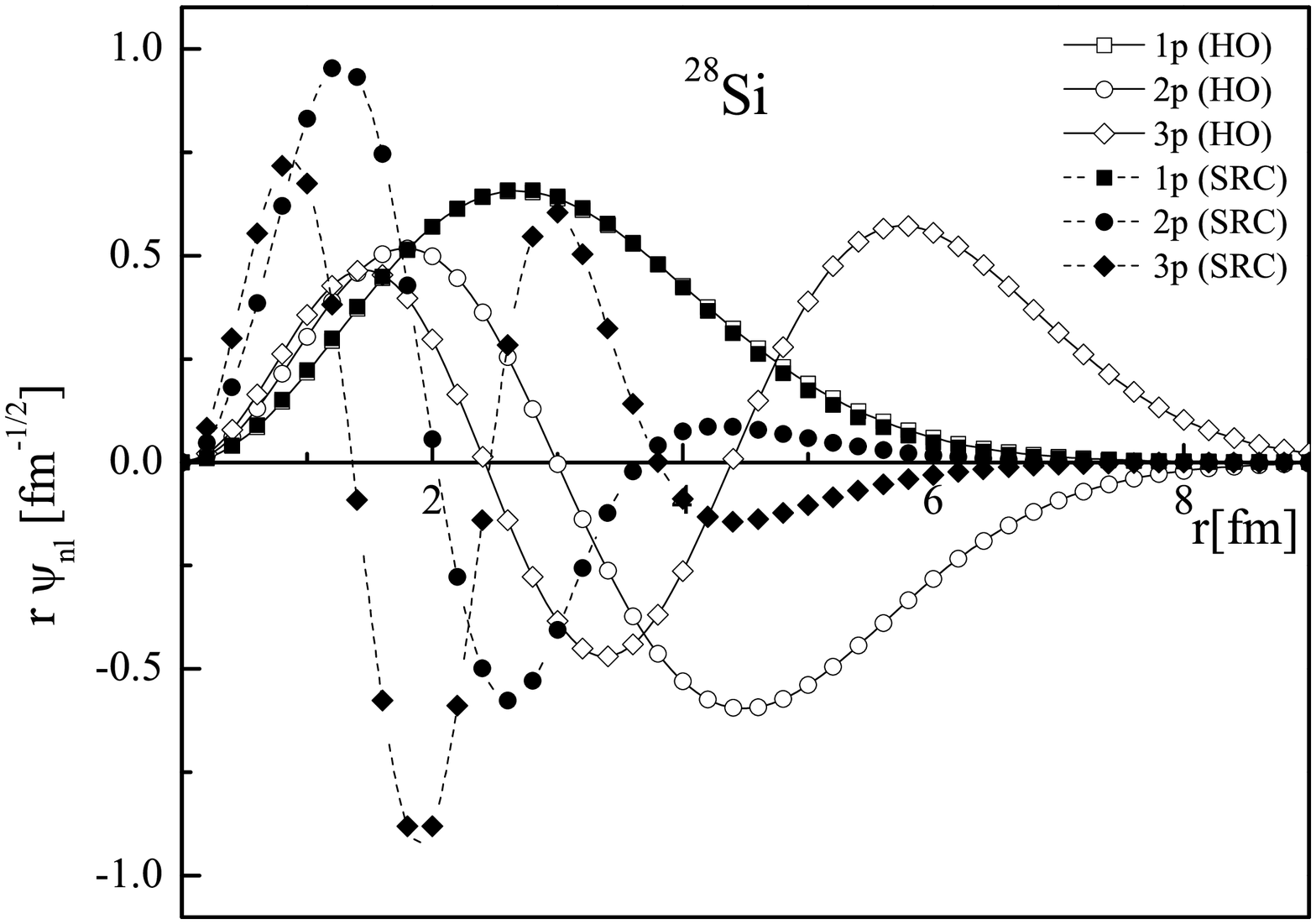}
    \hspace{-0.35in}
    \epsfxsize=2.75in
    \epsffile{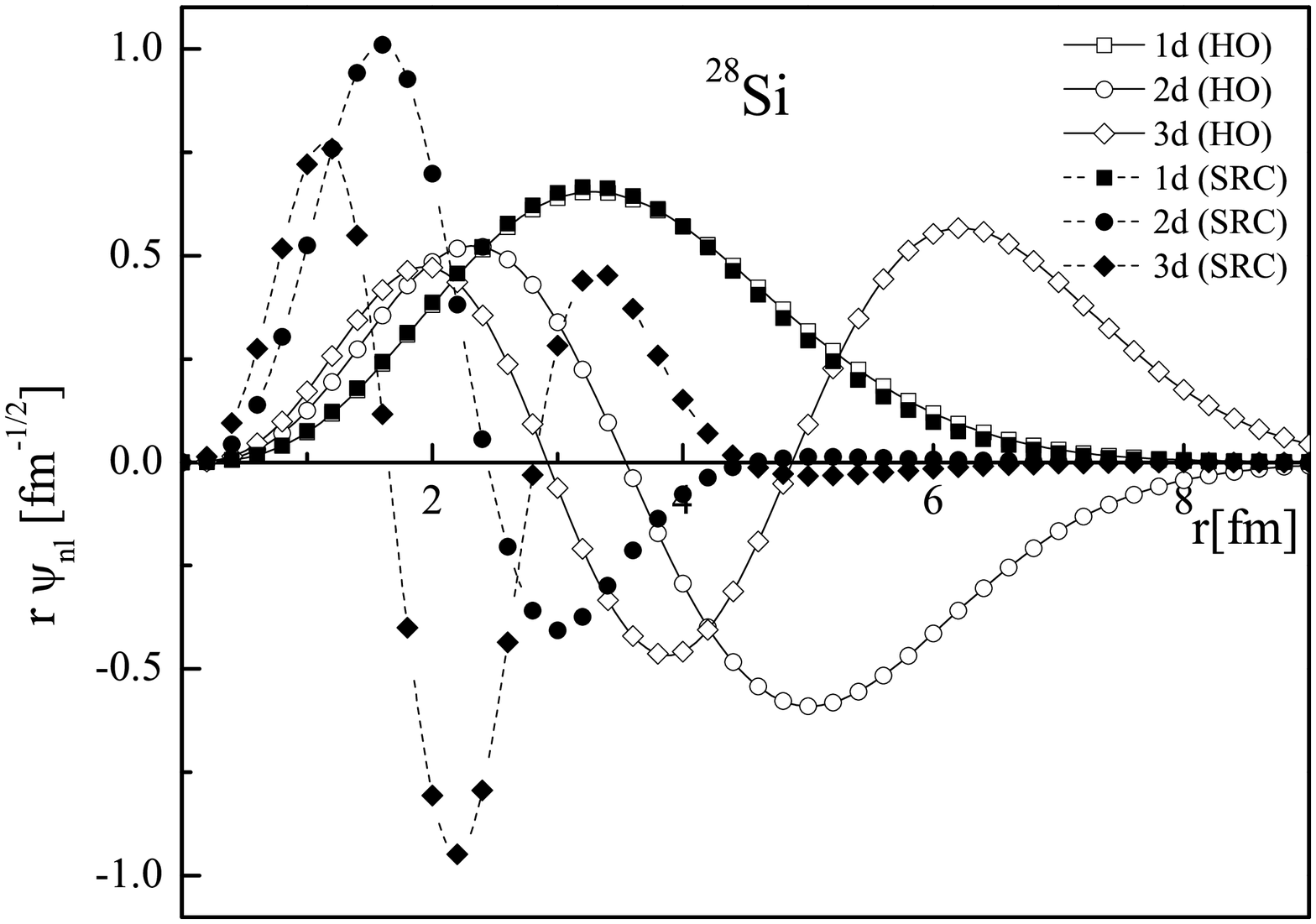}
    }
  }
  \vspace{-10px}
 \centerline{\hbox{ \hspace{0.0in}
    \epsfxsize=2.75in
    \epsffile{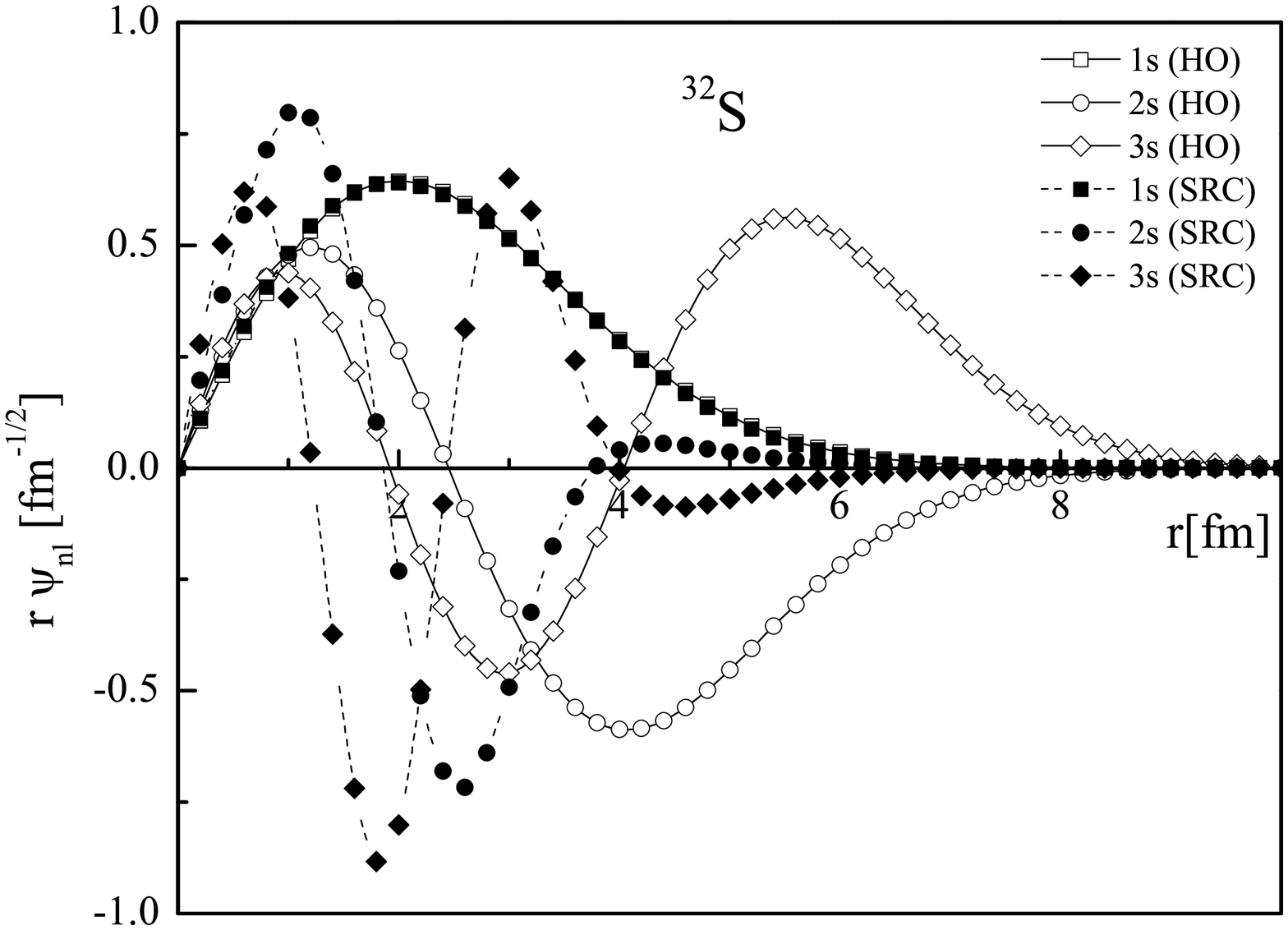}
    \hspace{-0.35in}
    \epsfxsize=2.75in
    \epsffile{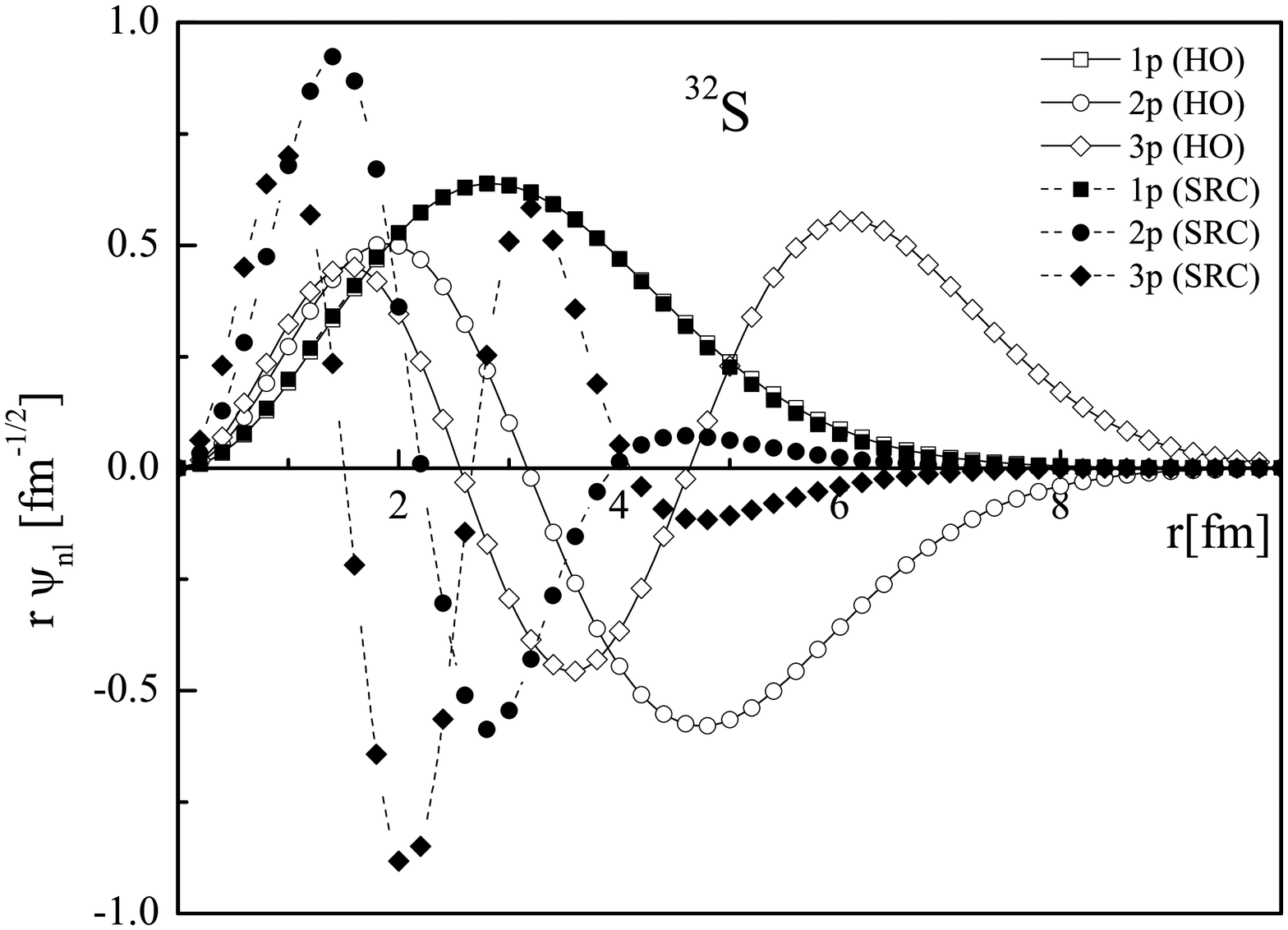}
    \hspace{-0.35in}
    \epsfxsize=2.75in
    \epsffile{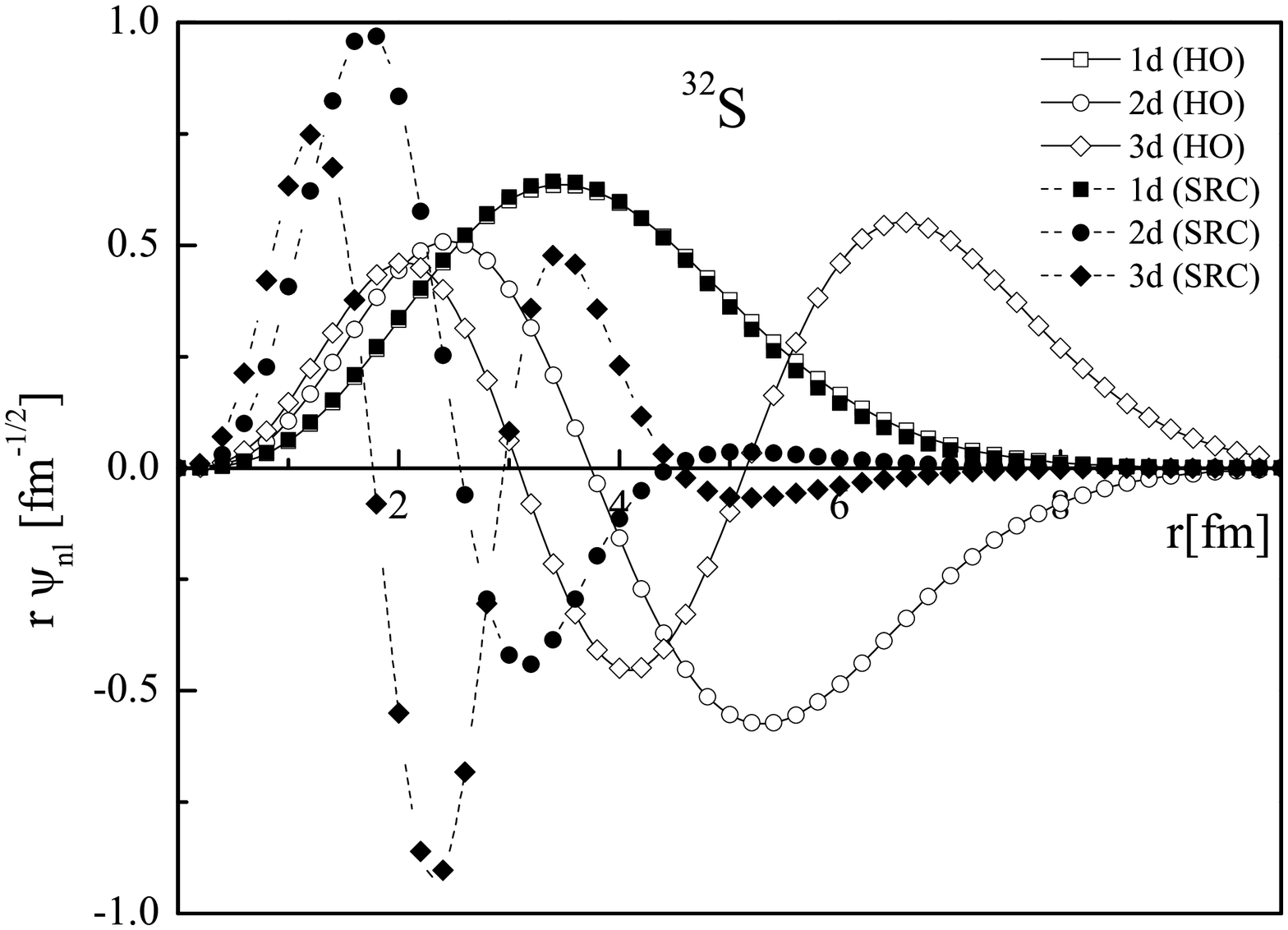}
    }
  }
\vspace{-1cm}
\caption{The natural orbitals of the particle- and hole-states (solid lines)
compared to the corresponding HO orbitals (dashed lines) for the
open shell nuclei  $^{12}$C, $^{24}$Mg, $^{28}$Si and $^{32}$S. }
\label{fg2}

\end{figure}

In Figure~\ref{fg1} and~\ref{fg2}, we compare the uncorrelated HO single-particle wave
functions with the corresponding correlated natural orbitals for
the considered nuclei. The most distinctive feature is the
similarity of the HO and the  NO single-particle wave functions
for the states lying below Fermi level. This is an indication that
the effect of short-range correlations on the configuration space
orbitals  is not significant for nucleon moving below Fermi level.
On the contrary, the main difference between the HO and the NO
wave functions appears for the states lying above the Fermi level.
Due to SRC, the particle NO wave functions are concentrated mostly inside
the nucleus while the HO ones are expanded outside the nucleus. As
a result the natural particle-state orbitals have smaller rms
radii and significantly larger high-momentum components compared
to the unoccupied MFA orbitals. In addition, they have  small but
non-zero occupation ratios. The above properties of the natural
orbitals affects appreciably the nucleon momentum distribution
$n(k)$. More precisely it was found that the high momentum
components of the  momentum distribution $n(k)$ can be
determined to a large extent by the particle-state contribution
due to SRC \cite{Stoitsov-93} .

\begin{figure}[hbtp]
\centerline{\hbox{ \hspace{0.0in}
    \epsfxsize=3.25in
    \epsffile{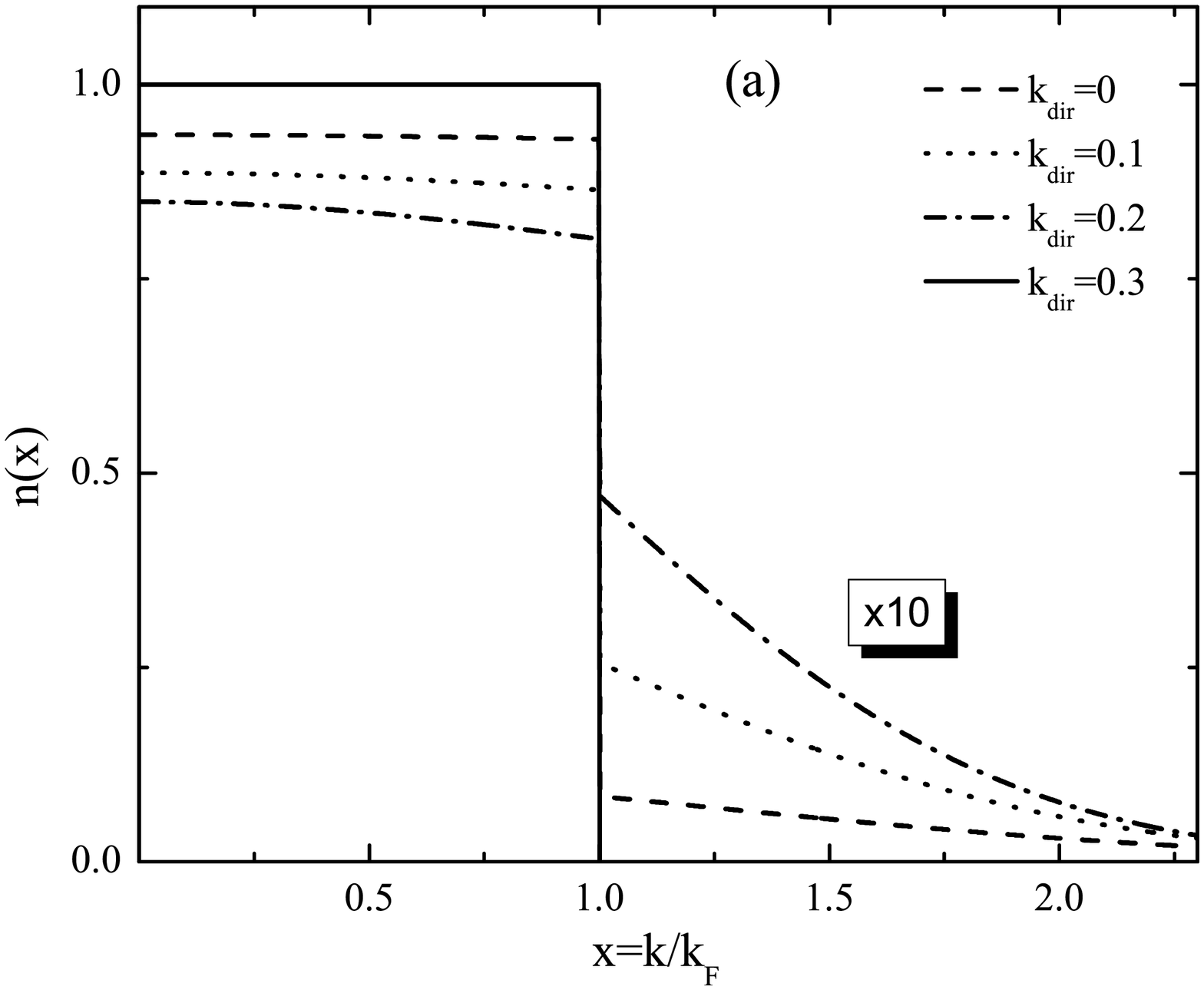}
    \hspace{-0.35in}
    \epsfxsize=3.25in
    \epsffile{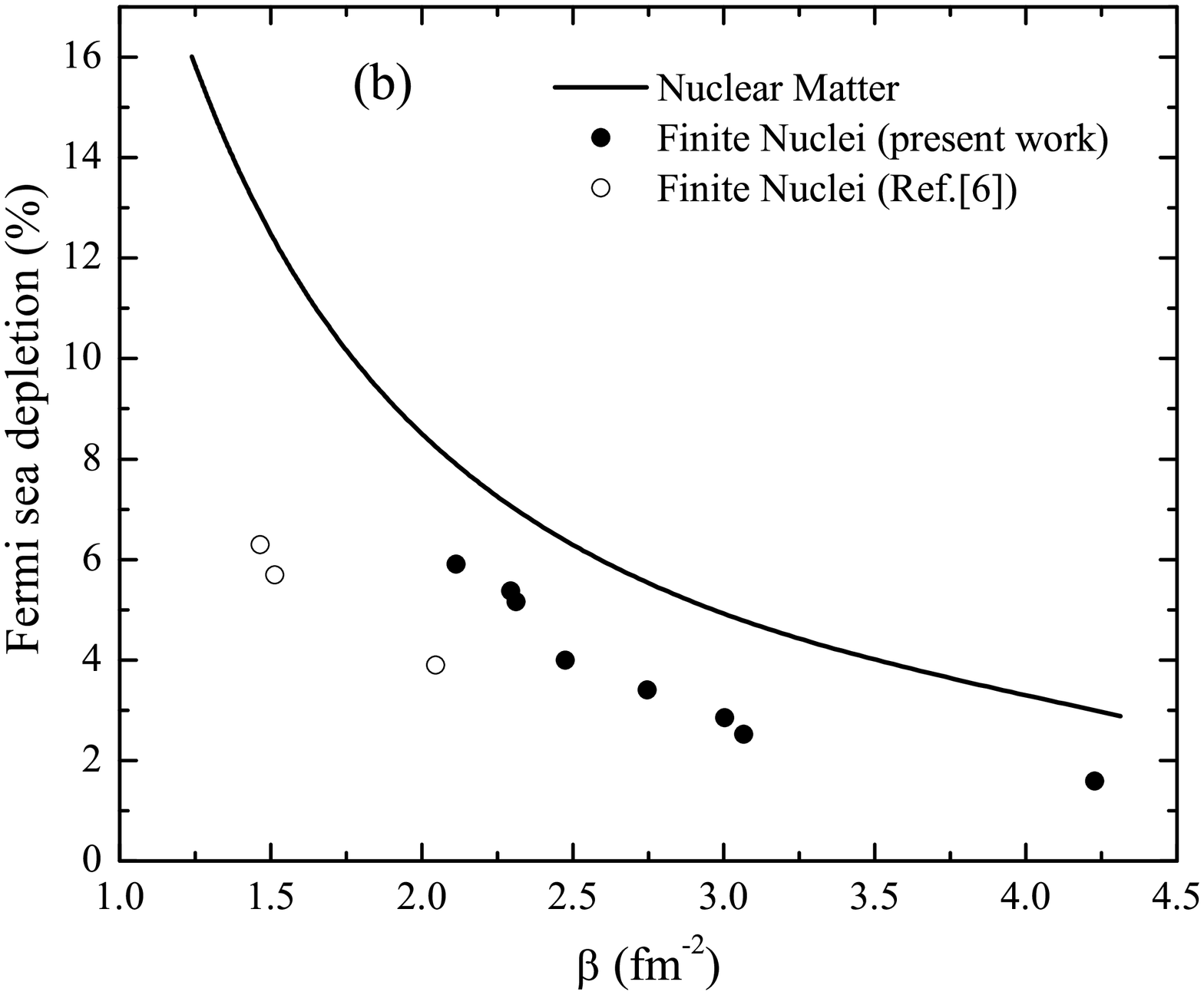}
    \vspace{-1cm}
     }
    }
        \vspace{-0.7cm}
\caption{(a)  The momentum distribution for correlated nuclear
matter versus $k/k_F$ for various values of the correlation
parameter $k_{\rm dir}$ (the values of $n(x)$ for $x > 1$ have been multiplied by 10) (b) The Fermi sea depletion of nuclear
matter (solid line), and finite nuclei of the present  work (solid
circles) and of Ref.~\cite{Stoitsov-93} (open circles).  } \label{fg3}
\end{figure}

The behaviour of the momentum distribution of the uniform nuclear
matter, by employing Eq.~(\ref{mn-mom-1}), as a function of
$k/k_F$ for various values of the wound parameter $k_{\rm dir}$ is
indicated in Figure~\ref{fg3}(a). The discontinuity $Z_F$, of the momentum
distribution at $k/k_F=1$, is another characteristic quantity
used as a measure of the strength of correlations of the uniform
Fermi systems and is defined as
\begin{equation}
Z_F=n(1^-)-n(1^+). \label{Z-F-1}
\end{equation}

The discontinuity $Z_F$ for various values of $k_{\rm dir}$ is displayed in Figure~\ref{fg3}(a).
$Z_F=1$ for the uncorrelated matter, while for
correlated matter $Z_F<1$. In the limit of very strong correlations
$Z_F=0$  there is no discontinuity on the momentum
distribution of the nuclear matter. The quantity ($1-Z_F$)
measures the ability of correlations to deplete the Fermi sea by
exciting particles from the hole states to the particle states
\cite{Flyn84}.

The Fermi sea depletion $D_{nm}$ in the case of uniform nuclear
matter is defined as
\begin{equation}
D_{nm}=\left(3\int_{1^{+}}^{\infty} n_{\rm LOA}(x)x^2
dx\right)\cdot 100\ (\%), \quad x=k/k_F .
\end{equation}

The depletion of the Fermi sea of finite nuclei and nuclear matter as a function of
the correlation parameter $\beta$ is shown in Figure~\ref{fg3}(b).
As a
comparison we also include the results of Ref.~\cite{Stoitsov-93}
where the OBDM of finite nuclei was evaluated using
Eq.~(\ref{OBDM-LOA}). This equation was employed in the present
work for the study of the Fermi sea depletion of nuclear matter.
In a similar work, Jaminon {\it et
al.} used the relativistic Brueckner-Hartree-Fock
approximation and they found that the FSD is around $5$\%
\cite{Mahaux-89}. In addition, de Jong {\it et al.}, in the
framework of a similar relativistic model found that the average
depletion of the Fermi sea is $11$\% \cite{Jong-91}.

It is obvious that the value of the  FSD, for the same value of the correlation
parameter $\beta$ is higher in nuclear matter compared to
finite nuclei. Actually, even for the same approximation for the
OBDM (that is the LOA  given by Eq.~(\ref{OBDM-LOA})) the
depletion in nuclear matter is almost twice compared to finite
nuclei. However it should be noted that the dependence of FSD on
$\beta$ exhibits almost similar behavior for nuclear matter and
finite nuclei. This is an indication that although the
finite nuclei have a different structure compared to
uniform nuclear matter, SRC affect their mean field structure
(uncorrelated case) in the same manner at least for nuclear matter
density close to the saturation density.

\section{Summary}
We studied the effect of short range correlations on the
natural orbitals and natural occupation numbers for various $sp$
and $sd$ shell nuclei via the correlated one-body density matrix.
A comparison with the MFA is also presented within the framework
of the Fermi sea depletion. It was found that  the effect of the
SRC on the natural orbitals  is not significant for nucleons
moving below the Fermi level. The main difference between HO and NO
wave functions appears for states lying above the Fermi level
affecting appreciably the high momentum components of the
momentum distribution $n(k)$.  Finally, we found that the value of the
Fermi sea depletion is higher in the closed shell nuclei
compared to the open shell nuclei but in both cases it is lower
compared to that of nuclear matter even for the same strength of SRC.
However definitive conclusions can only be obtained by future experimental investigation.

\section*{Acknowledgments}
Dr. Psonis is grateful for the A.U.Th Research Committee fellowship in support of this work.


\begin{table}[h]
\caption{The occupation ratio $\eta_a$ of particle- and hole-states for various $sp$ and $sd$ shell nuclei.
The last two columns correspond to theoretical and experimental data of Ref.~\cite{Stoitsov-93} and \cite{Kramer-90}, respectively.}
\label{tb1}
\vspace{5px}
{\begin{tabular}
 {c c c c c c c c c c c}
 \hline\hline
$nl$ & $^{4}$He  & $^{12}$C  & $^{16}$O &  $^{24}$Mg & $^{28}$Si & $^{32}$S & $^{36}$Ar & $^{40}$Ca   & $^{40}$Ca~\cite{Stoitsov-93} & $^{40}$Ca~\cite{Kramer-90} \\
\hline\
1s  & 0.9484 & 0.9044 & 0.8846 & 0.9497 & 0.9066 & 0.9200    & 0.8156   & 0.6630  &0.8899   & 0.9500\\
2s  & -0.0037 & 0.0029 & 0.0036 & 0.0009 & 0.0023 & 0.0019   & 0.0066   & 1.0686  &0.9588    & 0.8900\\
3s  & 0.0037 & -0.0017 & -0.0023 & 0.0005 & 0.0012 & 0.0011  & 0.0030   & 0.0089  &-0.0097   & \\
4s  & 0.0002 & 0.0009 & 0.0013 & 0.0003 & 0.0005 & 0.0005    & -0.0017   & -0.0089  &0.0113    & \\
\hline
1p  & 0.0077 & 0.6644 & 0.9852 & 0.9794 & 0.9564 & 0.9580    & 0.9021   & 0.8889   &0.9296  & 0.9200\\
2p  & 0.0008 & 0.0026 & 0.0031 & 0.0008 & 0.0020 & 0.0016    & 0.0053   & 0.0093   &-0.0055    & 0.1200\\
3p  & 0.0001 & 0.0007 & -0.0022 & 0.0004 & 0.0010 & 0.0009   & -0.0039   & -0.0054   & 0.0121       & \\
4p  & 0.0000 & -0.0005 & 0.0009 & 0.0002 & -0.0008 & -0.0005 & 0.0022   & 0.0037   &0.0041       & \\
\hline
1d  & 0.0025 & 0.0036 & 0.0052 & 0.4034 & 0.6050 & 0.8009    & 0.9988   & 1.0022    &0.9468    & 0.8920 \\
2d  & 0.0003 & 0.0011 & 0.0017 & 0.0008 & 0.0020 & 0.0016    & 0.0047   & 0.0066    &-0.0063      & \\
3d  & 0.0000 & 0.0003 & 0.0005 & 0.0004 & 0.0009 & 0.0008    & 0.0042   & -0.0052    &0.0082      & \\
4d  & 0.0000 & 0.0000 & 0.0002 & 0.0002 & 0.0003 & 0.0003    & 0.0018  & 0.0027   &0.0028        & \\
\hline
1f  & 0.0008 & 0.0020 & 0.0033 & 0.0009 & 0.0025 & 0.0024    & 0.0080   & 0.0090    & 0.0127     &  0.3700\\
2f  & 0.0001 & 0.0006 & 0.0010 & 0.0005 & 0.0011 & 0.0010    & 0.0031   & 0.0041    & 0.0050      &  \\
3f  & 0.0000 & 0.0002 & 0.0003 & 0.0002 & 0.0005 & 0.0005    & 0.0012   & 0.0017    & 0.0018      & \\
4f  & 0.0000 & 0.0000 & 0.0001 & 0.0001 & 0.0002 & 0.0002    & 0.0004   & 0.0007    & 0.0006       & \\
\hline
1g  & 0.0003 & 0.0013 & 0.0021 & 0.0007 & 0.0019 & 0.0017    & 0.0059   & 0.0065    &  0.0087      & \\
2g  & 0.0000 & 0.0004 & 0.0006 & 0.0003 & 0.0008 & 0.0008    & 0.0022   & 0.0028     & 0.0031             &  \\
3g  & 0.0000 & 0.0001 & 0.0002 & 0.0002 & 0.0003 & 0.0003    & 0.0008   & 0.0011     & 0.0010      &  \\
4g  & 0.0000 & 0.0000 & 0.0000 & 0.0001 & 0.0001 & 0.0002    & 0.0003   & 0.0004     &       &  \\
\hline\hline
\end{tabular}}
\end{table}

\begin{table}[h]
\caption{The depletion of the hole states   and the FSD (in $\%$) for various $sp$ and $sd$ nuclei compared to
the theoretical and experimental data of Ref.~\cite{Stoitsov-93} and \cite{Kramer-90}, respectively.}
\label{tb2}
\vspace{5px}
{\begin{tabular} {c c c c c c c c c c c} \hline\hline
$nl$ & $^{4}$He  & $^{12}$C  & $^{16}$O &  $^{24}$Mg & $^{28}$Si & $^{32}$S & $^{36}$Ar & $^{40}$Ca & $^{40}$Ca~\cite{Stoitsov-93}  & $^{40}$Ca~\cite{Kramer-90} \\
\hline
1s  & 5.16        & 9.56      & 11.54       & 5.03     & 9.34     & 8.02      & 18.44  & 33.7     &11.01   & 5.00\\
2s  &             &           &             &            &          &           &          & -6.86   & 4.12   & 11.00  \\
1p  &             & 0.34      & 1.48        & 2.06       &  4.36    & 4.20     & 9.79  & 11.11        &    7.04  &  8.00 \\
1d  &             &           &             & -0.85      & -0.83     & -0.11     & 0.12  & -0.22       & 5.32       & 10.80 \\
\hline
FSD  & 5.16            & 3.41          &  4.00           & 1.59     &   2.85   & 2.52     & 5.38  &  5.91      & 6.30       & 9.40 \\
\hline\hline
\end{tabular}}
\end{table}

\end{document}